\documentclass[twocolumn]{aa} 
\DeclareTextSymbol{\degre}{OT1}{23}

\usepackage{graphicx,natbib}
\usepackage{enumitem}
\usepackage{calc}
\usepackage{txfonts}
\usepackage{pifont}
\begin{document}

\title{From stellar nebula to planetesimals}

\authorrunning{Marboeuf et al.}

\author{Ulysse~Marboeuf\inst{1}, Amaury~Thiabaud \inst{1}, Yann Alibert \inst{1, 2}, Nahuel Cabral\inst{1}, \& Willy Benz \inst{1}}
\institute{${}^1$Physics Institute and Center for Space and Hability, University of Bern, Bern, Switzerland\\
${}^2$ Observatoire de Besan\c con, France\\
\email{ulysse.marboeuf@space.unibe.ch}}

\date{Received ??; accepted ??}

  
  \abstract
{Solar and extrasolar comets and extrasolar planets are the subject of numerous studies in order to determine their chemical composition and internal structure. In the case of planetesimals, their compositions are important as they govern in part the composition of future planets. 
}
{The present works aims at determining the chemical composition of icy planetesimals, believed to be similar to present day comets, formed in stellar systems of solar chemical composition. The main objective of this work is to provide valuable theoretical data on chemical composition for models of planetesimals and comets, and models of planet formation and evolution.}
{We have developed a model that calculates the composition of ices formed during the cooling of the stellar nebula. Coupled with a model of refractory element formation, it allows us to determine the chemical composition and mass ratio of ices to rocks in icy planetesimals throughout in the protoplanetary disc.}
{We provide relationships for ice line positions (for different volatile species) in the disc, and chemical compositions and mass ratios of ice relative to rock for icy planetesimals in stellar systems of solar chemical composition. From an initial homogeneous composition of the nebula, a wide variety of chemical compositions of planetesimals were produced as a function of the mass of the disc and distance to the star.
Ices incorporated in planetesimals are mainly composed of H$_2$O, CO, CO$_2$, CH$_3$OH, and NH$_3$.
The ice/rock mass ratio is equal to 1$\pm$0.5 in icy planetesimals following assumptions. This last value is in good agreement with observations of solar system comets, but remains lower than usual assumptions made in planet formation models, taking this ratio to be of 2-3.}
{}


\maketitle

\section{Introduction}
The determination of the chemical composition of planetesimals or (extrasolar) comets (assuming they have conserved the initial chemical composition of the protoplanetary disc since their formation) has been the subject of numerous studies. 
In the context of the solar system, comets are assumed to provide the link between the chemical composition of planets and that of the protosolar nebula.
These bodies are supposed to be the most primitive objects of stellar systems. They are formed from ices in cold areas of the protosolar nebula and should therefore have conserved the chemical composition of the cloud that gave birth to the stellar systems.
These small objects could be, at least partially, the origin of water and planetary atmospheres (Dauphas 2003) as a result of collision with planets during their migration towards the inner solar system (Ipatov \& Mather 2006, 2007). Moreover, it has been proposed that they could have taken part in the emergence of life on Earth.
The observations of the coma (gas phase) of comets in the solar system shows that these bodies can have chemical compositions that vary from one to two orders of magnitude from one comet to another (Bockel{\'e}e-Morvan et al. 2004; Mumma \& Charnley 2011).
The study of the chemical composition of comets, i.e. icy planetesimals, is of crucial importance to understanding these discrepencies as well as the process of planetary system formation and the chemical composition of planets. The study of the composition of volatile species incorporated in planets is an important parameter in planetary internal structure and thermal evolution models of planets (see Marboeuf et al. 2014 submitted, hereafter paper 3). \\
Our aim, in this paper and a companion one (Thiabaud et al. 2014, hereafter paper 1), is to determine the composition of planetesimals that are the first elements of future planets (see paper 3). We proceed in the following way. 
We first compute the composition of planetesimals for a population of different masses of discs, and at any distance to the central star, considering both the icy species (topic of this paper), and refractory species (paper 1). These results are then combined with the results of the formation model of Alibert et al. (2013) (which provides the amount of planetesimals accreted by every planet as a function of the distance to the central star) to derive the final composition of planets (see paper 1 for refractory elements and paper 3 for volatile species).
The main objective of the work presented in this paper is to provide valuable theoretical data of chemical composition for models of planetesimals and comets (in the context of the solar system), and for planet formation and evolution models. 

This article is organized as follows: in a first step (Sect.~\ref{models}), we present the models (both the planetary formation model, and the composition model) and describe the physical processes taken into account. In a second step (Sect.~3), we discuss the physical assumptions, chemical composition of the gas phase of the stellar nebula, and the adopted thermodynamics parameters. 
In a third step (Sect.~\ref{results_planetesimals}), we determine the chemical composition of grains incorporated in planetesimals formed at any distance to the star. We therefore compare our results with the chemical composition of comets in the solar system (Sect.~\ref{comparison_comets}). The last part (Sect.~\ref{conclusion}) is devoted to the discussion and conclusion.

\section{Description of the physical models used in this work \label{models}}

Our procedure to determine the chemical composition of planetesimals is organized as follows. 
Planetesimals are believed to form from the coagulation of small grains (icy and/or rocky), whose chemical composition is itself provided by the condensation/trapping sequence of gas in the discs.
It means that we follow the condensation sequence of the gas of the disc, and stop this sequence at a temperature which is the one in the disc at the initial time of the planet formation model (see paper 3).
This provides us with the composition of grains, which is assumed to be identical to that of planetesimals formed at the same location in the disc.
This approach is obviously simplified, since the chemical composition of planetesimals is the one resulting from the condensation/trapping sequence of gases at the location of the grains, and we do not include the radial drift of both small icy grains and planetesimals. 
We describe each part of this study below, providing physical assumptions.

\subsection{Model of disc accretion}

The model of disc evolution used in this work is described in Alibert et al. (2005, 2013). We provide hereafter a simple description of this model.
It allows us to determine the values of the temperature $T(r,z)$, pressure $P(r,z)$, and surface density $\Sigma(T,P,r,z)$ of the disc for different distances $r$ to the star (between 0.04 AU and 30 AU, i.e. the comet and planet formation zones), z-vertical axis, and for different turbulent viscosities $\nu(r, z)$ of the disc. The most important parameter for the evolution of $T(r,z)$, $P(r,z)$, and $\Sigma(T,P,r,z)$ is $\nu(r,z)$.
Following Shakura \& Sunyaev (1973), the turbulent viscosity $\nu$ is expressed in terms of Keplerian rotation frequency $\Omega$ and sound velocity $C_s$, as $\nu = \frac{\alpha C_s^2}{\Omega^2}$, where $\alpha$ is a parameter characterizing the turbulence in the disc.

The model computes the vertical structure at a given radial distance $r$ from the star by solving a system of three equations describing the hydrostatic equilibrium, the energy conservation, and the radiative flux (see Sect. 2.1.1 in Alibert et al. 2005). 
By integrating the pressure, temperature, and surface density on the $z$-axis of each radial distance $r$ from the disc surface $H$ to the midplane disc, this provides the radial temperature $T(r)$, pressure $P(r)$, and surface density $\Sigma(T,P,r)$ in the midplane of the disc for each distance $r$ to the star.

At the beginning of the planet formation process (see paper 3, and Alibert et al. 2005, 2013), the gas disc surface density profile $\Sigma(r)$ is given by using the following relation (see Eq.~(20) in Alibert et al. 2013),
\begin{equation}
\Sigma(r) = \Sigma_0 (\frac{r}{a_0})^{-\gamma}  exp\left[-\left(\frac{r}{a_{core}}\right)^{2-\gamma}\right] \qquad (g \, cm^{-2}),
\end{equation}
where $r$ is the distance to the star (in AU), $a_0$ is equal to 5.2 AU, $a_{core}$ is the characteristic scaling radius (in AU), $\gamma$ is the density slope, and $\Sigma_0$ (g.cm$^{-2}$) the surface density of the disc at 5.2 AU defined as
\begin{equation}
\Sigma_0 =  (2-\gamma) \frac{M_{disc}}{2\pi a_{core}^{2-\gamma} a_0^\gamma}  \qquad (g \, cm^{-2}).
\end{equation}
The values for $a_{core}$, $\gamma$, and $M_{disc}$ vary for every disc (see Table~4 in paper 1). \\
The knowledge of $\Sigma(r)$ allows us to determine the pressure $P(r)$ and temperature $T(r)$ in the disc at the end of condensation/trapping process. 

We use two approaches. The first considers that the disc is not irradiated (hereafter non-irradiated model). In this case, the temperature of the surface of the disc is $T_s = T_b$, where $T_b$ is the background temperature, and only the viscous accretion determines the thermodynamic properties of the disc. The second considers that the disc is fully irradiated by the star (hereafter irradiated model).
In this case, both the radiation and viscous accretion determine the thermodynamic properties of the disc. The effect of the irradiation by the central star is modelled by modifying the temperature boundary condition at the disc surface (see Fouchet et al. 2012). This temperature is set to $T^4_s = T^4_b + T^4_{s,~ir}$, where the irradiation temperature $T_{s,~ir}$ is derived from Hueso \& Guillot (2005),
\begin{equation}
T_{s,~ir} = T_* \left[\frac{2}{3\pi} \left(\frac{R_*}{r}\right)^3  +  \frac{1}{2} \left(\frac{R_*}{r}\right)^2 \left(\frac{H_p}{r}\right) \left(\frac{d~ln~H_p}{d~ln~r} - 1\right)\right]^{1/4}   \qquad (K),
\end{equation}
where $R_*$ is the stellar radius and $H_P$ is the pressure scale height (see Fouchet et al. 2012 for more details). The first term on the right corresponds to the flux that would be intercepted by a flat disc, the second term is the contribution of the flaring of the disc to the stellar flux absorption.

We note that this irradiated model does not take into account all physical processes such as the absorption and scattering effects of the incident flux, and the absorpbion and re-radiation of the flux in the disc. Therefore, the irradiation effect at the mid-plane is maximum for this model (see Fouchet et al. 2012).
Moreover, if the effect of grain growth and settling towards the disc mid-plane were taken into account, the temperature should decrease (e.g. D'Alessio et al. 2001; Dullemond \& Dominik 2004). In addition, possible shadowing effect could work and the dust temperature should decrease. Ideally, all these effects should be taken into account in the disc but the irradiated (unrealistically hot) and non-irradiated (unrealistically cold) models represent two extreme cases and allow us to frame the thermodynamic conditions of ice formation in the cooling disc and migration of planets (see paper 3) for the two extremes of hot and cold conditions.

\subsection{Model of gas condensation and gas trapping \label{model_gas_condensation}}

The chemical composition of volatile molecules incorporated in icy grains and later in icy planetesimals, is determined by both the process of condensation and the process of gas trapping in water ice (hereafter clathrate formation) on the surface of refractory grains during the cooling of the stellar nebula.
Clathrates are crystalline solids composed of water and gas that can be formed during the cooling of the stellar nebula only if water molecules are already condensed at higher temperature ($\approx$ 150-170 K). The water molecule structure is then organized in the form of cages which are stabilized by the inclusion of gas molecules ($\neq$ H$_2$O). Each cage contains a single gas molecule trapped because of van der Waals interactions. 

\begin{figure}[h]
\begin{center}
\includegraphics[width=7.cm, angle=-90]{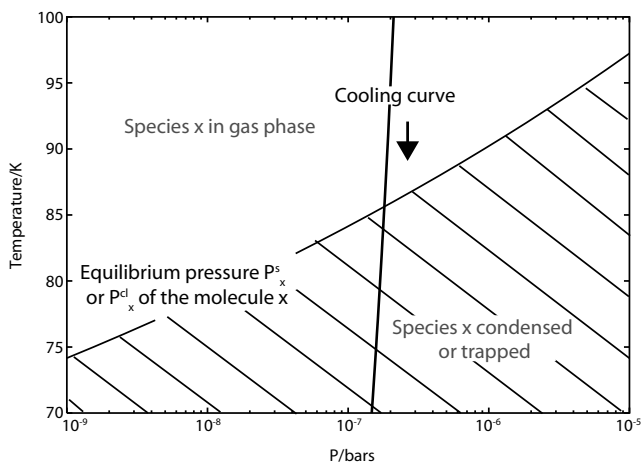}
\caption{Schematic view of the process of gas condensation or trapping in the disc. The nearly horizontal line is the stability curve of pure condensate or clathrate of the species $x$. The nearly vertical line is the cooling curve of the protoplanetary disc (the evolution of the disc proceeds from high to low temperatures). Above the stability, the gas species remains in the gas phase of the disc. Below, the dashed area corresponds to the area of stability of ices or clathrates (area of gas condensation or trapping in water ice).}
\label{process_condensation}
\end{center}
\end{figure}

The thermodynamic conditions of gas condensation/trapping (resp. ice/clathrate formation) are determined by the partial pressure of volatile molecules in the gas phase, and the temperature of the disc. The process by which volatile molecules are condensed and/or trapped in clathrates is described as follows (see fig.~\ref{process_condensation}). At a given distance $r$ from the star, the total pressure $P^{tot}(r)$ of the gas phase in the midplane of the disc, its temperature $T(r)$, and its surface density $\Sigma(T,P,r)$ decrease with time. 
By decreasing, the partial pressure $P_x$ of the species $x$ in the gas phase can become lower than the equilibrium pressure $P^s_x$ of condensation (respectively $P^{cl}_x$ for clathrate formation). At this time, the species $x$ is assumed to be condensed (respectively trapped) at the surface of the refractory grains in the disc.
The mass ratio $\Gamma_x^g(r)$ of the species $x$ relative to H$_2$O (the first volatile molecule to condense in the disc) in icy grains (and therefore, following our assumptions, in icy planetesimals), at the distance $r$ from the star can be determined by the relation given by Mousis \& Gautier (2004),
\begin{equation}
\Gamma_x^g(r) =\frac{Y_x}{Y_{H_2O}} \frac{\Sigma(T_x,P_x,r)}{\Sigma(T_{H_2O},P_{H_2O},r)},
\end{equation}
where $Y_x$ and $Y_{H_2O}$ are, respectively, the mass ratio of the volatile molecules $x$ and H$_2$O relative to H$_2$ in the protoplanetary disc during their condensation/trapping at the surface of refractory grains; $\Sigma(T_x,P_x,r)$ and $\Sigma(T_{H_2O},P_{H_2O},r)$ are, respectively, the surface densities of the disc at $r$ when the volatile molecules $x$ ($\neq$ of H$_2$O) and H$_2$O are condensed or trapped.
A given species $x$ is never condensed (respectively trapped) on the surface of refractory grains as long as its partial pressure $P_x$ remains greater than the equilibrium pressure $P^s_x$ of condensation (respectively $P^{cl}_x$ for clathrate formation) at a given distance $r$ to the star. In this case, the volatile molecule will not be subsequently incorporated in planetesimals.
This situation always happens near the star where the temperature/pressure is always larger than the temperature/pressure of condensation of H$_2$O ($\approx$ 150-170 K).

In the case of clathrate formation, another condition must be fulfilled: the process  stops when no more crystalline water ice is available to trap gases. Indeed, the trapping of gases in cages of clathrate structure requires a large quantity of water (around six molecules of water are used in order to capture one molecule of gas; see the hydrate number $n^{hyd}$ defined in Sect.~\ref{clathrate_formation}). It can therefore happen that a species $x$ is only partially trapped, the rest being condensed (at lower temperatures) as pure ice as soon as the disc becomes sufficiently cold with time. 
We note that if a planetesimal does not contain H$_2$O molecules, other volatile molecules are not incorporated in these objects ($\Gamma_x(r)$  = 0) as they condense or are trapped at lower temperature during the cooling of the stellar nebula.

\section[]{Physical assumptions and parameters adopted in models \label{parameter}}
\subsection[]{Species considered in the stellar nebula \label{secparam1}}

In this paper, we focus the study on the volatile components of the stellar nebula. We assume that all volatile molecules are composed of H, O, C, N, and S atoms in solar abundances (Lodders 2003). The gas-phase composition is determined by taking into account both refractory and volatile components. Refractory elements include both mineral and organic elements (see paper 1 for details). 

The presence and origin of complex organic components in protostellar regions and their possible incorporation into protoplanetary discs, i.e. in planetesimals, are active topics of research (e.g. Pollack et al. 1994; Greenberg et al. 1995; Dartois 2005; Boogert et al. 2008; Oberg et al. 2008). 
Organic refractory materials are disequilibrium species that can be produced by extensive processing of ices in specific environments such as low-temperature, UV-irradiated environments of the ISM, cosmic rays (Pollack et al. 1994; Greenberg et al. 1995; Boogert et al. 2008). Among the large organic molecules observed or suspected in diffuse clouds, one finds polycyclic aromatic hydrocarbons (PAHs), fullerenes, carbon chains, diamonds, amorphous carbon (hydrogenated and bare), and complex kerogen-type aromatic networks.
These elements, less volatile than ices, condense at temperatures lower than 500 K (a typical evaporation temperature of organics; see Greenberg \& d'Hendecourt 1985; Kouchi et al. 2002; Schutte \& Khanna 2003; Boogert et al. 2008) and can trap H$_2$O molecules at high temperatures before their condensation (Boogert et al. 2008) during the cooling of the stellar nebula. 

Complex organic refractory components have also been discovered in comets and meteorites. The spacecraft investigations of the comet 1P/Halley revealed the presence of organic refractory grains (Kissel et al. 1986a,b; Jessberger et al. 1988) known today as CHON-PS for their carbon, hydrogen, oxygen, nitrogen, phosphorus and sulfur content. From Halley observations, Jessberger et al. (1988) determined the relative proportion of atoms C, O, and N included in refractory organic compounds. The carbonaceous chondrites, incorporated in meteorites, contain a small percent of carbon and some of them exhibit a large variety of organic compounds (Cronin et al. 1993).

From these observations and studies, it is natural to search if organic compounds, observed in small bodies of the solar system, have been formed in ISM or during the cooling of the solar nebula from gas volatile molecules, and could have been incorporated unaltered into comets, or if they have been formed from species condensed (or trapped) as ices (or refractories), and subsequently formed within these objects during their thermal evolution around the Sun (Mumma \& Charnley 2011).

The answer is currently not trivial and we decide in this paper to study two extremes cases:\\ 
$\bullet$ 1) model with refractory organic - The first considers that organic compounds formed in ISM
remain unaltered during the collapse of the molecular cloud and the cooling of the stellar nebula. In this case, we assume that refractory organics are initially present in the gas phase of the protoplanetary disc \footnote{The origin of the organic refractory material is probably not the same for each of these species. One can, for example, imagine that species with a low number of C atoms may condense in the gas phase, whereas other species with more C atoms are formed later on.}.
The O, C, N, and S atoms are shared between rocks, refractory organic and volatile molecules in ISM as follow: 26\% of the total oxygen in the nebula is devoted to rocks (Lodders 2003);
the fractional abundance of organic carbon is assumed to be 60\% $\pm$ 15\% of the total carbon in molecular clouds (Pollack et al. 1994; Sekine et al. 2005); and the relative proportion of C:O:N included in refractory organics is supposed to be 1:0.5:0.12 (Jessberger et al. 1988; Pollack et al. 1994; Sekine et al. 2005).
The remaining of atoms O, C, N, and S are part of volatile molecules.\\
$\bullet$ 2) model without refractory organic - The second case considers that refractory organics compounds are destroyed during the collapse of the molecular cloud. Atoms O, C, N, and S are shared only between rocks and volatile molecules in ISM, organic compounds being formed during the thermal evolution of ices in the solar system. In this last case, only 26\% of the total oxygen in the nebula is devoted to rocks (Lodders 2003). The remaining atoms O, C, N, and S are devoted to volatile molecules.
Table~\ref{paramdiscatoms} provides solar abundances of the O, C, N, and S atoms relative to H$_2$ and their repartition between rocks, organics, and volatile molecules for both models. For refractory components, the reader is referred to paper 1.

\subsection[]{Choice of chemical species and their abundances in ISM \label{secparam2}}

The ISM is composed mostly of hydrogen and helium gas (74 \% and 25 \% in mass, respectively) with a trace of heavier elements (1 \% in mass) (White 2010).
At the beginning of the computation, we assume that heavier elements are composed of a mixture of volatile molecules such as H$_2$O, CO, CO$_2$, CH$_4$, H$_2$S, N$_2$, NH$_3$, and CH$_3$OH. Except N$_2$, these molecules are the most abundant volatile species  observed in the ISM (Gibb et al. 2000; van Dishoeck 2004; Gibb et al. 2004; Whittet et al. 2007; Boogert et al. 2011; Mumma \& Charnley 2011) and in solar cometary comas (Bockel\'ee-Morvan et al. 2004; Crovisier 2006; Mumma \& Charnley 2011). Hereafter, we discuss the abundances of each volatile molecule.\\
H$_2$O is the most common interstellar ice component around refractory grains in molecular clouds (Gibb et al. 2000; Boogert \& Ehrenfreund 2004; Dartois 2009).
Consequently, the abundances of other volatile molecules refer to H$_2$O on a relative scale in ISM\footnote{This is also the case for comets of the solar system (Bockel\'ee-Morvan et al. 2004; Crovisier 2006)} (Dartois 2005; Mumma \& Charnley 2011). Note hereafter that the relative abundance of species (to H$_2$O) are given in \% of mol of H$_2$O.
CO$_2$ is the second most common ice component with an abundance varying between 10\% and 50\% relative to H$_2$O (Gibb et al. 2004; Bennett et al. 2009; Dartois 2009), the majority of sources having an abundance in the range of 20 - 30\% (Pontoppidan et al. 2008). 

CO ice is largely observed in molecular clouds. Depending on the density and temperature of clouds, it is partially distributed between the gas phase of the molecular cloud and the ices at the surface of refractory grains (Dartois 1998; Gibb et al. 2004). However, except in very hot or shocked regions (Boonman et al. 2003; Nomura \& Millar 2004; Lahuis et al. 2007; Pontoppidan et al. 2008), this molecule remains mainly in the gas phase of molecular clouds (Dartois 2009), while the CO$_2$ and H$_2$O molecules are mainly condensed at the surface of refractory grains (Bergin et al. 1995). Hence, the gas to ice ratios of H$_2$O and CO$_2$ can reach, respectively, 0.18 and 0.04, although although the CO ratio can reach a factor of 3-4 (Dartois 1998). The abundance of CO ice varies among sources, from 3\% to 20 - 30\% relative to H$_2$O (Boogert \& Ehrenfreund 2004; Gibb et al. 2004; Bennett et al. 2009; Dartois 2009). Hence, the total abundance of CO relative to H$_2$O can reach 100\% (see Table 3 in Mumma \& Charnley 2011).

CH$_3$OH abundance varies by an order of magnitude (Dartois et al. 1999; Pontoppidan et al. 2003; Boogert \& Ehrenfreund 2004; Mumma \& Charnley 2011) in molecular clouds (variations from 1 to 30\%).
This species can be dominant in some areas of molecular clouds, becoming the second most abundant species after H$_2$O, and is at the same time almost absent in other locations, a variation whose origin is still debated (Dartois 2009; Mumma \& Charnley 2011).
In contrast to CH$_3$OH, CH$_4$ has similar abundance in all the sources for which data are available (Gibb et al. 2000).
Indeed, its abundance in cometary ices shows some similarities with interstellar and protostellar ices (Gibb et al. 2000). 
Moreover, for a large sample of low-mass YSOs, the solid CH$_4$ abundance with respect to H$_2$O is centred at 5.8\% with a standard deviation of 2.7\% (in the range 2-8\%) in the sources (Oberg et al. 2008).\\

While H$_2$O, CO, CO$_2$, CH$_3$OH, and CH$_4$ seem to be accurately characterized, there are still many unknowns regarding the nitrogen component of the icy mantle around refractory grains (Gibb et al. 2004). 
NH$_3$ abundance can vary between 2\% and 15\% relative to H$_2$O (Mumma \& Charnley 2011).
Since N is an abundant cosmic element and N$_2$ is a stable molecule, the latter should be present in ISM, but its abundance remains difficult to constrain because it is inactive in the infrared (Dartois 2005). Here, we adopt a N$_2$:NH$_3$ ratio of 1:1, a value compatible with thermo-chemical models of the solar nebula (Lewis \& Prinn 1980).
S is assumed to exist only in the form of H$_2$S. The abundance of this molecule varies by an order of magnitude between 0.3 and 4 for quiescent dense clouds and massive protostars (Mumma \& Charnley 2011).\\

Given the chemical abundances observed in ISM, we assume that the molar ratio H$_2$O:CH$_3$OH:CH$_4$:CO$_2$ is equal to 100:15:6:20 and that the ratio of H$_2$O:N$_2$:NH$_3$:H$_2$S is equal to 100:7:7:2. The two most abundant carbon volatile compounds CO and CO$_2$ have a CO:CO$_2$ ice ratio ranging from about 1 to 4 in all the sources, including quiescent and low/intermediate-mass regions (Gibb et al. 2004). Considering the contributions of CO in both ice and gas phases, we also adopt the CO:CO$_2$ ratio of 1:1 and 5:1 in our model as minimum (hereafter poor CO model) and maximum (hereafter rich CO model) values. These values are consistent with the ISM measurements considering the contributions of both gas and ice phases (Mumma \& Charnley 2011). This leads to H$_2$O:CO ratios of 5:1 (resp. CO:CO$_2$=1:1) and 1:1 (resp. CO:CO$_2$=5:1).

Knowing the abundance of the volatile molecules CH$_3$OH, CH$_4$, CO$_2$, CO, N$_2$, NH$_3$, and H$_2$S relative to H$_2$O (see Table \ref{paramdisc}), and the repartition of the O, C, N, and S atoms between rocks, refractory organics, and gas species in the ISM (see Table \ref{paramdiscatoms}), we can infer the amount of H$_2$O by using the equation
\begin{equation}
H_2O \approx \frac{\xi^v}{\chi};
\label{equa_conservation}
\end{equation}
here $\chi$ is defined as
\begin{equation}
 \chi = 1+\frac{CO}{H_2O} + \frac{CO_2}{H_2O} + \frac{CH_3OH}{H_2O} + \frac{CH_4}{H_2O} + \frac{NH_3}{H_2O} + \frac{N_2}{H_2O} + \frac{H_2S}{H_2O}, 
\end{equation}
where the ratios of molar abundances of volatile molecules $x$ to H$_2$O (X:H$_2$O) and to all the ices (X:Ices) are given in Table \ref{paramdisc};
$\xi^v$ is the sum of the abundances of O, C, N, and S atoms incorporated in volatile molecules and defined as
\begin{equation}
\xi^v=\xi - (\xi^m + \xi^o),
\end{equation}
where $\xi$, $\xi^m$, and $\xi^o$ are the sum of the abundances of the O, C, N, and S atoms in solar nebula (including volatile molecules, refractory organic compounds and minerals), in minerals and in refractory organic compounds, respectively. The repartition of atoms between volatile molecules, organics, and refractory elements is given in Table \ref{paramdiscatoms}.

\begin{table*}
\centering 
\caption{Initial molar abundances of atoms in the stellar nebula for all models}
\begin{tabular}{l|c||ccc|ccc}
\hline
\hline
Models						&																	& \multicolumn {3} {c|}{with refractory organics}	&  \multicolumn {3} {c}{without refractory organics}		\\
\hline								
Species X         &	Solar	(X$_{ISM}$)$^{*}$								& \multicolumn {6} {c}{X/X$_{ISM}$}\\		
\hline		
      &												&	  ices (X$^{ices}_{ISM}$) & organics & minerals &ices (X$^{ices}_{ISM}$) & organics & minerals\\		 	
		O	& 1.16 10$^{-3}$					&	0.59 & 0.15 & 0.26	& 0.74 & 0 & 0.26	\\
		C	& 5.82 10$^{-4}$					&	0.4 &	0.6 &	0	 & 	1.0 & 0 &	0	\\
		N	& 1.6 10$^{-4}$					  &	0.74 &	0.26	&	 0 & 1.0 &	0 &	 0		\\
		S	& 3.66 10$^{-5}$					&	0.5 & 0	&  0.5 & 0.5 & 0	&  0.5		\\
		\hline
		C:O & 0.5                   & 0.68 & 4 & 0 &  1.36 & 0 & 0 \\ 
		N:C & 0.275									&	1.85 & 0.43 & 0 & 0.275 & 0 &  0 \\
\hline
\end{tabular}
\begin{flushleft}
$^{*}$Data provided by Lodders (2003).
\end{flushleft}
\label{paramdiscatoms}
\end{table*}

For both rich (CO:CO$_2$=5:1) and poor (CO:CO$_2$=1:1) CO models, the amount of H$_2$O in ices represents respectively 56\% and 39\% (in mol) of ices (the sum of H$_2$O, CO, CO$_2$, CH$_3$OH, CH$_4$, N$_2$, NH$_3$, and H$_2$S) incorporated in grains. The increase in the carbon species CO induces the decrease in H$_2$O in ices. This leads to differences in the abundances of each volatile component (relative to ices). For the first assumed chemical composition (CO:CO$_2$=1:1), CO and CO$_2$ represent the second most abundant molecules with an abundance of 11\% (in mol) relative to ices for each of them. For the second assumed chemical composition (CO:CO$_2$=5:1), CO and CO$_2$ represent respectively 39\% and $\approx$8\% (in mol) of all volatile molecules.

\begin{figure}[h]
\begin{center}
\includegraphics[width=9.cm, angle=0]{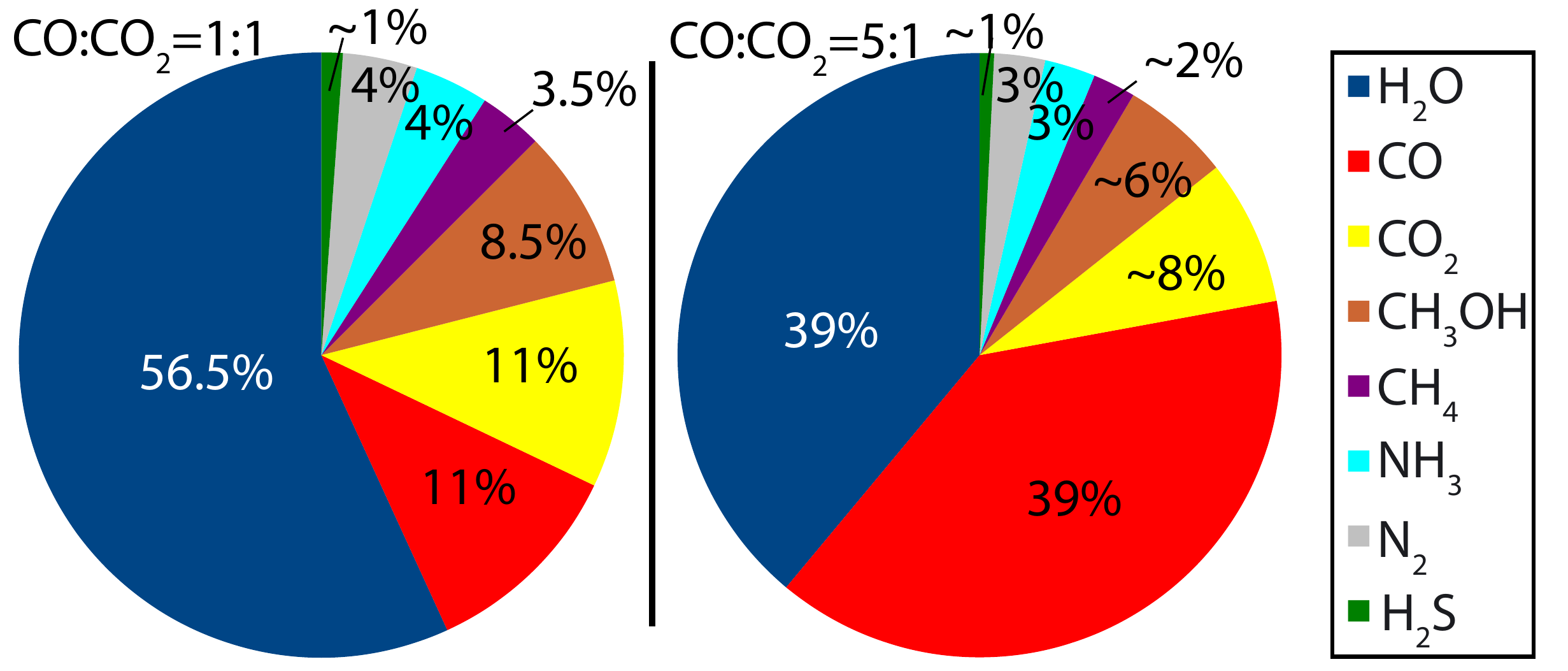}
\caption{Schematic view of molar abundances of species H$_2$O, CO, CO$_2$, CH$_3$OH, CH$_4$, N$_2$, NH$_3$, and H$_2$S relative to ices (the sum of these species) in the molecular cloud (ISM). Model with CO:CO$_2$ $=$ 1:1 and 5:1.}
\label{fig_abundances}
\end{center}
\end{figure}

Table \ref{paramdisc} provides the molar abundance of species $x$ in the molecular cloud for both models (CO:CO$_2$=1:1 and 5:1). For clarity, we have provided abundances relative to H$_2$O, to ices, and to H$_2$. To illustrate the chemical composition of the molecular cloud, Fig.~\ref{fig_abundances} presents the molar abundance of species H$_2$O, CO, CO$_2$, CH$_3$OH, CH$_4$, N$_2$, NH$_3$, and H$_2$S relative to ices for both poor and rich CO models.

\begin{table*}
\centering 
\caption{Initial molar abundances of volatile molecules in the stellar nebula for all models}
\begin{tabular}{l|ccc||ccc}
\hline
\hline
Models &  \multicolumn {3} {c||}{CO:CO$_2$ = 1:1} & \multicolumn {3} {c}{CO:CO$_2$ = 5:1} \\
\hline
Molecules X					& X/H$_2$O (\%) & X/Ices$^{*}$ (\%) &	X/H$_2$ 			&  X/H$_2$O (\%) &  X/Ices$^{*}$ (\%) &  X/H$_2$  \\
H$_2$O	&		100	 &   56.5 &		5.95	10$^{-4}$   			  &	  100		& 39 		&	4.1 10$^{-4}$\\ 
CO	    &		20	 &   11	  &   1.19 10$^{-4}$      			&		100 	&	39 		&	4.1 10$^{-4}$	\\
CO$_2$	&	  20	 &	 11   &   1.19 10$^{-4}$	  		  	&		 20		&	7.78 	&	8.19 10$^{-5}$	\\
CH$_4$	&	   6	 &   3.4  &   3.57 10$^{-5}$	          &	    6		&	2.3 	&	2.46 10$^{-5}$		\\
H$_2$S	&	   2	 &   1.13 &	1.19	10$^{-5}$	            &		  2  	& 0.778	& 8.19 10$^{-6}$		\\
N$_2$	  &		 7	 &   3.95 &	4.16 10$^{-5}$	            &	    7		& 2.72	&	2.87 10$^{-5}$	\\
NH$_3$	&	   7	 &   3.95 &	4.16	10$^{-5}$	            &	    7		& 2.72	&	2.87 10$^{-5}$	\\
CH$_3$OH&	  15	 & 	 8.47 &	8.92 10$^{-5}$	            &	   15		& 5.84	&	6.14 10$^{-5}$	\\
\hline
\end{tabular}
\begin{flushleft}
$^{*}$Ices= H$_2$O + CO + CO$_2$ + CH$_4$ +  H$_2$S +  N$_2$ +  NH$_3$ +  CH$_3$OH
\end{flushleft}
\label{paramdisc}
\end{table*}

\subsection[]{Gas phase chemistry in the disc \label{gaschemistry}}

Protoplanetary discs are the intermediate step between the cloud collapse stage and the planetary system stage. The chemical and physical properties of discs provide the initial conditions for planet formation. Both radiation from the central star and viscous accretion are the main sources of energy, dominating disc physics and chemistry (Aresu et al. 2012).
In this study, we assume that the temperature of the stellar nebula (after the collapse of the molecular cloud) is high enough everywhere, and for all initial masses of discs, to sublimate all ISM ices within 0 - 30 AU, the zone of formation of comets and planets. Thus, similarly to the case of the solar nebula (Mousis et al. 2002), the gas-phase abundance of volatile molecules in the disc derives directly from the abundance in the ISM discussed in Sect.~\ref{secparam2}.\\
This assumption raises the question of how the volatile molecules originally formed in ISM can survive in a hot and energetic protostellar nebula chemistry.
Hereafter, we discuss some of the possible chemical processes that can modify the relative abundance of species in the disc.
Indeed, the ISM material could undergo chemical modification during the collapse and accretion phase of protostellar evolution (Mumma \& Charnley 2011).
Depending upon where and when this material enters the disc, any interstellar signatures may be completely lost or largely retained (Mumma \& Charnley 2011; Neufeld \& Hollenbach 1994; Aikawa et al. 2008; Lee et al. 2008; Charnley \& Rodgers 2009; Visser et al. 2009; Visser \& Dullemond 2010).\\

Visser et al. (2011) offer a general picture of the chemical history of the different zones of the disc.
Material that ends up in different parts of the disc encounters different physical conditions during the collapse and therefore undergoes a different chemical evolution.
Physico-chemically, discs can be divided into three layers: a cold zone near the midplane, a warm molecular layer at intermediate altitudes, and a photon-dominated region at the surface (Bergin et al. 2007; Visser et al. 2011).
Surface of protoplanetary discs are irradiated by UV and X-ray photons from the central star as well as UV photons and cosmic rays originating from the interstellar medium. These radiation play an important role in the chemistry and thermal balance of the disc (Aresu et al. 2012). Molecules and atoms are altered at the surface and in the intermediate layers of discs by chemical and physical processes such as heating and thermochemical reactions, UV photochemistry, X-ray ionization, neutral and ion-molecule chemistry, and physical mixing or dilution with/by other nebular materials. As the radiation is attenuated through optical depth and/or distance, the gas becomes cooler in the midplane (colder region) of the disc.
In this last area, molecules are probably altered by gas-grain chemistry.
However, as discussed by Ciesla \& Charnley (2006), the mechanisms, reaction rates, and pathways in many of the proposed interstellar schemes are highly uncertain. Hence, detailed conclusions derived from disc chemistry models that are strongly dependent on surface chemistry should be regarded with caution (Ciesla \& Charnley 2006). In this work, we assume that the gas-grain chemistry does not occur during and after the cooling of the disc.

The total H$_2$O, CO, and N$_2$ abundances (by considering gas and ice phases) do not change significantly during the post-collapse phase, while the total CH$_4$ and NH$_3$ abundances change by more than two orders of magnitude in a large part of the disc (Visser et al. 2011). There are two areas in the disc where the abundances of all species remain nearly constant: near the surface out to ~ 10 AU, and at the midplane between 5 AU and 25 AU. The chemistry in the first area is dominated by fast photoprocesses. The second area, near the midplane, is the densest part of the disc and therefore has high collision frequencies. In both cases, the chemical timescales are short and the chemistry reaches equilibrium during the final stages of accretion (Visser et al. 2011).

The main carbon reservoir is CO. This molecule is very stable and its chemistry is straightforward. The bulk of the CO is situated in the optically thick part of the disc, and the total CO gas mass does not change when varying the FUV (far-ultraviolet) and X-ray radiation field (Meijerink et al. 2012). CO does not undergo any processing except the evaporation process in the collapse phase of the nebula (Visser et al. 2011). During the remaining part of the infall trajectory of volatile species in the disc, the main carbon species are all largely converted into CO, excepted CH$_4$ whose abundance is well coupled with CO in the disc (Visser et al. 2011).
The only major chemical event for this molecule is the evaporation during the initial warm-up of the disc. It is never photodissociated or never reacts significantly with other species (Visser et al. 2011). The distribution of CO$_2$ is similar to that of CO in the inner disc, existing only in the midplane with a maximum value of 10$^{-4}$ within a few AU of the star (Walsh et al. 2010). Beyond this radius, its abundance decreases significantly. H$_2$O is not stable to both FUV and X-rays (Meijerink et al. 2012) at the surface of protoplanetary discs. This molecule evaporates and is photodissociated. However, as shown in Visser et al. (2011), this area represents less than 7\% of the total mass of the disc, and the total (by considering gas and ice phases) H$_2$O abundance does not change significantly.
Prinn \& Fegley (1981, 1989), Lewis \& Prinn (1980), Mousis et al. (2002), and Talbi \& Herbst (2002) determined that the amount of carbon species CO, CO$_2$, and CH$_4$ produced and lost through chemical reactions in a gas dominated by H$_2$ (see Eqs.~\ref{eq1} and \ref{eq2} below) within protoplanetary discs is negligible, except at distances quite close to the star, from which no ices finally incorporated in planets originate :
\begin{equation}
$CO$ + $H$_2$O$ \rightleftharpoons $CO$_2 + $H$_2,
\label{eq1}
\end{equation}
\begin{equation}
$CO$ + 3 $H$_2 \rightleftharpoons $CH$_4 + $H$_2$O$.
\label{eq2}
\end{equation}
The authors found that the initial CO:CO$_2$ and CO:CH$_4$ ratio in the disc are little affected throughout the solar nebula, except quite close to the star.
As a result, the amount of carbon species produced through this reaction is not significant during the whole lifetime of the protoplanetary discs.

The evolution of N$_2$ in the disc parallels that of CO, because they have similar binding energies and are both very stable molecules (Bisschop et al. 2006); they show no significant change in abundance across the inner 30 AU of the disc below the surface photodissociation layer. 
The evolution of NH$_3$ shows a lot more variation (Visser et al. 2011). This molecule does not survive in the surface layers where it is quickly destroyed by reaction with abundant CN, producing HCN. NH$_3$ only exists in regions shielded from UV radiation (Willacy and Woods 2009). 
The study of the chemical reaction linking N$_2$ and NH$_3$ in a gas dominated by H$_2$ in the disc (see reaction Eq.~\ref{eq3} below and Lewis \& Prinn 1980 for details) shows that, due to kinetic effects, a reduction of the abundance of N$_2$ of more than 1\% is unlikely over the lifetime of the nebula :
\begin{equation}
$N$_2 + 3 $H$_2 \rightarrow 2 $NH$_3.
\label{eq3}
\end{equation}
As a result, we assume that the abundances of nitrogen species adopted in the ISM remains the same in the gas phase of the disc.

The surface density of the stellar nebula in the model still remains higher than 50 g.cm$^{-2}$ in the area of condensation/trapping of gas species at the surface of grains (in the midplane of the disc). So, we assume in our study that changes of abundances of species due to FUV and X-ray radiation field are negligible.

To summarize, the aforementioned studies show that chemical reactions are not likely to change the abundance of volatile molecules in the disc, except for some species such as NH$_3$ and CH$_3$OH.
Note also that all the volatile molecules (H$_2$O, CO$_2$, CH$_4$, H$_2$S, NH$_3$, and CH$_3$OH) except CO and N$_2$ are easily destroyed in the gas-phase by reactions with ions near surface layers of the disc in 10 AU (see Visser et al. 2011), and does not keep high abundances once they are evaporated from grains (Doty et al. 2002; Nomura \& Millar 2004). However, grain surface reactions in the disc will recover the loss, and the abundances of volatile molecules in ice may not be very different from those adopted in this study.
We also assume that the abundance of all species in the gas phase of the midplane in the disc (therefore prior to the condensation), are the same as the ones in the ISM. We note that this assumption is consistent with the chemical composition of comets in the solar system, which is qualitatively and, in many cases, quantitatively consistent with the one determined for the major components of astronomical ices (see e.g. Mumma 1997; Irvine et al. 2000; Langer et al. 2000; Gibb 2000; Mumma \& Charnley 2011). 

\subsection[]{Thermodynamics parameters of condensation and clathrate formation \label{clathrate_formation}}

The model for computation of the chemical composition of ices formed during the cooling of the stellar nebula takes into account both processes of condensation (ice formation) and trapping (clathrate formation) of volatile molecules $x$.

The thermodynamic conditions for condensation of volatile molecules $x$ ($\neq$ of H$_2$O) at the surface of refractory grains are given by the saturation equilibrium pressure $P_x^s$,
\begin{equation}
ln(P_x^s)= \sum_{i=0}^5{\frac{A_i}{T^i}},
\label{equasublim}
\end{equation}
where $P_x^s$ is expressed in bar and $T$ in K; $A_i$ are constants and their values are given in Table \ref{thermoparam}. These values are taken from the compilation of Fray \& Schmitt (2009).\\
For H$_2$O, the saturation equilibrium pressure adopted is
\begin{equation}
ln~\left(\frac{P_{H_2O}^s}{P_t}\right) = \frac{3}{2} ln~\left(\frac{T}{T_t}\right) + \left(1- \frac{T_t}{T}\right)~\eta\left(\frac{T}{T_t} \right),
\label{equaH2O1}
\end{equation}
where $T_t$ (273.16K) and $P_t$ (6.11657.10$^{-3}$ bar) are, respectively, the temperature and pressure at the triple point for H$_2$O and $\eta$ expressed as
\begin{equation}
\eta\left(\frac{T}{T_t} \right) = \sum_{i=0}^6 e_i\left( \frac{T}{T_t}\right)^i,
\label{equaH2O2}
\end{equation}
where the $e_i$ coefficients are given in Table~\ref{thermoparamH2O}.

\begin{table*}
\centering 
\caption{Parameters of the equilibrium curves of condensation of ices on refractory grains used in Eq.~(\ref{equasublim}).}
\begin{tabular}{l|c|cccccc}
\hline
\hline
 Molecules &  Valid temperature range for values  &  $A_0$	  &   $A_1$	 &   $A_2$	&   $A_3$	  &  $A_4$	  &  $A_5$ \\
 \hline
CO  &      $T$ $\leq$ 61.55 K                       & 10.43   &  -721.3     & -1.074 10$^4$ & 2.341 10$^5$ & -2.392 10$^6$ & 9.478 10$^6$  \\
   &  		61.55 K $\leq$ $T$ $\leq$ 68.1 K	        	   &  10.25     &  -748.2  &-5.843 10$^3$ &3.939 10$^4$  & 0& 0 \\

CO$_2$ &    $T$ $\leq$ 194.7 K             & 14.76 & -2571 & -7.781 10$^4$ & 4.325 10$^6$ & -1.207 10$^8$ & 1.35 10$^9$  \\
 &     		194.7 K $\leq$ $T$ $\leq$ 216.58 K	& 18.61 & -4154 & 1.041 10$^5$  &   0           & 0             &0         \\
 CH$_4$  & 	20.6 K $\leq$ $T$ $\leq$ 90.68 K	&    10.51   & - 1110 & -4341       & 1.035 10$^5$ & -7. 91 10$^5$ &   0           \\
H$_2$S    &     $T$ $\leq$ 126.2 K        & 12.98 & -2707 &  0   &   0   &   0   &   0     \\
  &  		126.2 K $\leq$ $T$ $\leq$ 187.57 K		&  8.933 &  -726 &  -3.504 10$^5$ & 2.724 10$^7$ & -8.582 10$^8$ & 0  \\
  N$_2$ & 10 K $\leq$ $T$ $\leq$ 35.61 K & 12.4 &	-807.4 &	-3.926 10$^3$	& 6.297 10$^4$ &	-4.633 10$^5$	 & 1.325 10$^6$ \\	
        & 35.61 K $\leq$ $T$ $\leq$ 63.14 K &  8.514  & -4.584 10$^2$ &	-1.987 10$^4$ &	4.8  10$^5$	&	-4.524 10$^6$ & 0\\
 NH$_3$ & 15 K $\leq$ $T$ $\leq$ 195.41 K & 15.96  &  -3.537  10$^3$ & 	-3.31 10$^4$ & 	1.742 10$^6$ &	-2.995  10$^7$ & 0  \\
 CH$_3$OH &  $T$ $\leq$ 157.36 K & 19.18 & 	-5.648  10$^3$ & 0&0&0&0\\
          & 157.36 K $\leq$ $T$ $\leq$ 175.5 K & 17.06 & 	-5.314  10$^3$ & 0&0&0&0\\
\hline
\end{tabular}
\label{thermoparam}
\tablefoot{All the sublimation/condensation equilibrium data are taken from Fray \& Schmitt (2009).}
\end{table*}

\begin{table}
\centering 
\caption{Parameters of the equilibrium curves of condensation of H$_2$O on refractory grains used in Equations (\ref{equaH2O1}) and (\ref{equaH2O2}).}
\begin{tabular}{lc}
\hline
i  & e \\
\hline
0		&	20.9969665107897 \\
1   & 3.72437478271362 \\
2   & -13.9205483215524 \\
3   & 29.6988765013566 \\
4   & -40.1972392635944 \\
5   & 29.7880481050215 \\
6   & -9.13050963547721 \\
\hline
\end{tabular}
\label{thermoparamH2O}
\tablefoot{The sublimation/condensation equilibrium data are taken from Fray \& Schmitt (2009).}
\end{table}

The formation of clathrates at the surface of icy grains occurs by gas-water ice interaction. This process can generate a chemical composition of icy planetesimals that differ significantly from models without clathrate formation since this structure selectively traps some gases  (see papers of Van der Waals \& Platteeuw 1959; Parrish \& Prausnitz 1972; Lunine \& Stevenson 1985; Kang et al. 2001; Sun \& Duan 2005; Anderson 2007; Rydzy et al. 2007; Thomas et al. 2009; Mousis et al. 2010). 
Only five volatile molecules considered in this work (CO, CH$_4$, CO$_2$, H$_2$S, and N$_2$) can be trapped in cages. To our best knowledge, no experimental data concerning the stability curve of the CH$_3$OH clathrate has been reported in the literature and the conditions under which it forms stoichiometric hydrates or clathrate (Blake et al. 1991; Notesco \& Bar-Nun 2000) is still unclear. NH$_3$ is known to form rather stoichiometric hemihydrates (2NH$_3$ - H$_2$O) and/or monohydrates (NH$_3$ - H$_2$O) under some conditions (Lewis 1972; Bertie \& Shehata 1984; Lunine \& Stevenson 1987; Kargel 1998; Moore et al. 2007). This molecule has also been assigned a role of anti-freeze for water ice and clathrate formation, modifying the stability region of the solid ice and methane clathrate hydrate phases as a thermodynamic inhibitor (Choukroun et al. 2006; Fortes \& Choukroun 2010; Choukroun et al. 2010; Shin et al. 2012). 
However, one study has shown that ammonia can form clathrates and participate synergistically in the formation of the cages in the presence of methane gas at low temperature (Shin et al. 2012), but the conditions under which clathrates are formed remains unclear and no experimental data allowing us to determine the conditions of formation of the NH$_3$ clathrate has been reported in the literature. So, in this work, it is assumed that NH$_3$ does not form a clathrate hydrate but rather a stoichiometric monohydrate (NH$_3$-H$_2$O).

Clathrates can exist mainly under two types of structure called structure I (SI) and structure II (SII) that contain two types of cavities (small and large) whose number, size, and hydrate number $n^{hyd}$ (mean number of water molecule per molecule of trapped species) differ and depend on the size of the trapped molecules (see Sloan \& Koh 2008; Gabitto \& Tsouris 2010 for a review).
The hydrate number $n^{hyd}$ therefore provides the average number of H$_2$O molecules needed to form cages around each gas species $x$ trapped,
\begin{equation}
n^{hyd} = \frac{N^{hyd}_{H_2O}}{N^{hyd}_x},
\label{hyd}
\end{equation}
where $N^{hyd}_{H_2O}$ is the number of water molecules forming the bulk clathrate structure and $N^{hyd}_x$ the number of gas molecules trapped inside.
Four of these five molecules (CO, CH$_4$, CO$_2$, H$_2$S) form individually a SI structure of clathrate (Davidson et al. 1987; Dartois 2011; Handa 1986b; Anderson 2003; Miller 1961) and N$_2$ is the only one molecule forming a SII-structure (Chazallon \& Kuhs 2002; Sloan \& Koh 2008), leading to small differences in $n^{hyd}$. All volatile molecules studied in this work are small enough to enter both the small and large cavities (Sloan \& Koh 2008; Gabitto \& Tsouris 2010). The ideal hydrate number n$^{hyd}$ in the case of a maximum occupancy of the cages (all cages are filled) is equal to 5.75 (resp. 5.67) for SI (resp. SII) structures (Sloan \& Koh 2008; Gabitto \& Tsouris 2010).
However, all cages are not necessarily filled: experimental data (Handa 1986a, 1986b; Sloan \& Koh 2008; Sun \& Mohanty 2006) have shown that the real values of n$^{hyd}$ are closer to 6 with some empty cavities. Since it is currently impossible to determine the real occupancy of the cages in the thermodynamic conditions of the cooling of the stellar nebula, we therefore fix the hydrate number $n^{hyd}$ to 6 whatever the gas molecule trapped and clathrate type structure (SI or SII).

The physical conditions for the formation of cages of clathrates by trapping gas species $x$ are given by the equilibrium pressure $P_x^{cl}$,
\begin{equation}
ln(P_x^{cl})= A_0 + \frac{A_1}{T^1},
\label{equaclat}
\end{equation}
where $P_x^{cl}$ is expressed in Pa and $T$ in K. $A_0$ and $A_1$ are constant and their values are given in Table \ref{clatpressure}.

Considering only the thermodynamic conditions of clathrate formation implies that the kinetics of their formation is fast enough compared to the cooling time of the disc.
The kinetic conditions of clathrate formation from crystalline water ice and gas species $x$ are, however, poorly constrained at low temperature and very few values exist only for CO$_2$ and CH$_4$ molecules and for temperatures greater than 180~K (Schmitt 1986; Staykova et al. 2003, 2004; Kuhs et al. 2006; Wang et al. 2002; Genov \& Kuhs 2003; Falenty et al. 2011). The physical parameters that may affect the kinetics are multiple: temperature, total pressure of the gas, activity of the water ice surface (mobility of water molecules), type of volatile molecule trapped, thickness of clathrate formed, and thermal history (Schmitt 1986).
Since it is currently impossible to determine the kinetic feasibility of the formation of clathrates in the thermodynamic conditions of the cooling of the stellar nebula, we examine two extreme processes of ice formation: the first assumes that clathrate formation is not possible and all volatile molecules condense only as pure ices on the surface of refractory grains; the second assumes that clathrate formation occurs during the cooling of the stellar nebula following the thermodynamic conditions. In the second case, we consider that all water molecules of the ice mantle around refractory grains can interact with volatile molecules of the gas phase in the stellar nebula to form clathrates. The clathration process stops therefore when no more H$_2$O molecules of the crystalline ice are available to trap volatile species $x$.

\begin{table}
\centering 
\caption{Parameters of the equilibrium curves of single-guest clathrates used in Eq.~(\ref{equaclat}).}
\begin{tabular}{l|cc|l}
\hline
\hline
Molecules			&					$A_0$	  &   $A_1$	 		& References (Exp Data)  \\	
 \hline
CO    &		23.3818    	 & 	-1890.39 &  \footnotesize{Mohammadi \& Richon (2010)$^{*}$}\\
CO$_2$ &     23.294         &	-2565  &  \small{Fray et al. (2010)}\\
CH$_4$  &  22.6864    &   -2176.2 &    \small{Fray et al. (2010)} \\
H$_2$S  &  22.8931    &  -3111.02 &   \small{Hersant et al. (2004)} \\
N$_2$   &  22.719     &  -1696.05 &  \small{Mohammadi \& Richon (2010)$^{*}$}  \\
NH$_3$  &   19.515     &  -2878.23&   \small{Hersant et al. (2004)}   \\
CH$_3$OH &   \multicolumn {2} {c|}{no data} & - \\
\hline
\end{tabular}
\begin{flushleft}
$^{*}$The empirical correlation laws for CO and N$_2$ clathrates are determined by fitting the data of Mohammadi \& Richon (2010).
\end{flushleft}
\label{clatpressure}
\end{table}

Once formed, we assume that the composition of ice is not altered by gas-grain chemistry in the disc (Willacy \& Woods 2009); in general, all of the physical changes that could occur during the post-cooling phase are not taken into account in this study.

\section{Chemical composition of icy planetesimals \label{results_planetesimals}}

\begin{figure}[h]
\begin{center}
\includegraphics[width=9.cm]{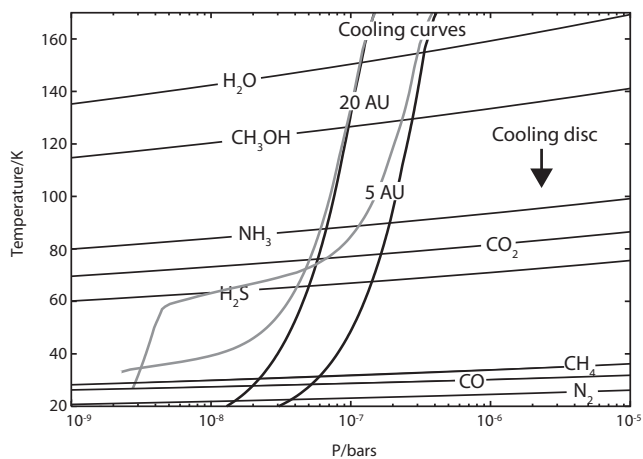}
\caption{Stability curves of pure condensates (horizontal lines) and cooling curves of the protoplanetary disc model (vertical lines) at two given distances to the star: 5 and 20 AU (the evolution of the disc proceeds from high to low temperatures), for irradiated (grey lines) and non-irradiated (black lines) discs. Species remain in the gas phase above the stability curves. Below, they condense as pure ices. Model with CO:CO$_2$ = 1:1.}
\label{cooling_condensation}
\end{center}
\end{figure}

We emphasize that we only study the abundances of volatile molecules incorporated in planetesimals. For refractory components, the reader is referred to paper 1. We apply 16 different initial physico-chemical conditions to several types of discs in order to determine all the possible physico-chemical compositions of planetesimals: we vary the chemical composition of the disc (see Table~\ref{paramdisc}), and study extreme assumptions on physical processes and thermodynamic conditions by taking into account, or not, the formation of clathrates, refractory organics, and the radiation in the disc.
We focus on the chemical composition of ices incorporated in planetesimals as a function of the distance to the star. The process by which volatile molecules are condensed/trapped in icy planetesimals is illustrated in figures \ref{cooling_condensation} and \ref{cooling_clathrates}. It is calculated using the stability curves of hydrates, clathrates, and pure condensates (horizontal lines), and the evolutionary tracks (vertical lines) giving the evolution of temperature and pressure of the protoplanetary disc between 0.04 AU and 30 AU (see sects.~\ref{model_gas_condensation} and 3.4 for details), for irradiated (grey lines) and non-irradiated (black lines) discs. The drop in the cooling curve for the irradiated disc is due to the decrease in the surface density with time: it corresponds to the transition between a thick disc and a thin one. The thermodynamic conditions of the condensation and clathrate formation of volatile molecules are determined by the intersection of the stability curve (horizontal lines) with the thermodynamic path of the nebula (cooling  curve, vertical lines) calculated at a given distance to the star, from 0.04 AU to 30 AU : a species is condensed/trapped when its equilibrium pressure $P^s_x$/$P^{cl}_x$ is lower than or equal to its partial pressure (total pressure of the disc times the abundance of the species in the disc) in the gas phase of the disc.

Whatever the chemical composition and model adopted, H$_2$O is the first volatile molecule to condense in the disc. 
For models without clathrate formation, we assume that H$_2$O molecules are not free to form cages of clathrates, or, equivalently, that the formation of clathrate is kinetically unfavourable. 
In this case, other volatile molecules condense as pure ices as soon as the temperature of the disc is low enough (see Fig.~\ref{cooling_condensation}) at a given distance $r$\footnote{Remember that the final temperature of the disc is known from the calculation of the final surface density profile $\Sigma(r)$ by using Eq.~(1)}. After H$_2$O, chemical species condense in the following order: CH$_3$OH, NH$_3$, CO$_2$, H$_2$S, CH$_4$, CO, and N$_2$ as soon as the disc is cool enough at the position $r$.

\begin{figure}[h]
\begin{center}
\includegraphics[width=9.cm]{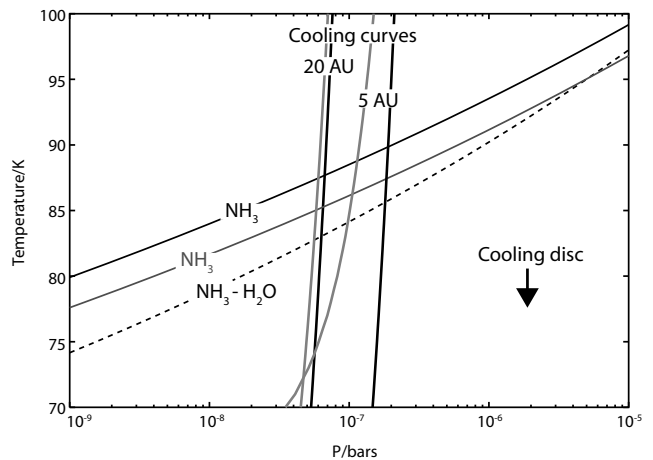}
\caption{Stability curves of pure condensates (solid lines) and hydrates of NH$_3$ (dashed line). The equilibrium pressure derived from old relation (see text) is in grey. The relation from Fray \& Schmitt (2009) is in black (solid line).}
\label{NH3}
\end{center}
\end{figure}

\begin{figure}[h]
\begin{center}
\includegraphics[width=9.cm]{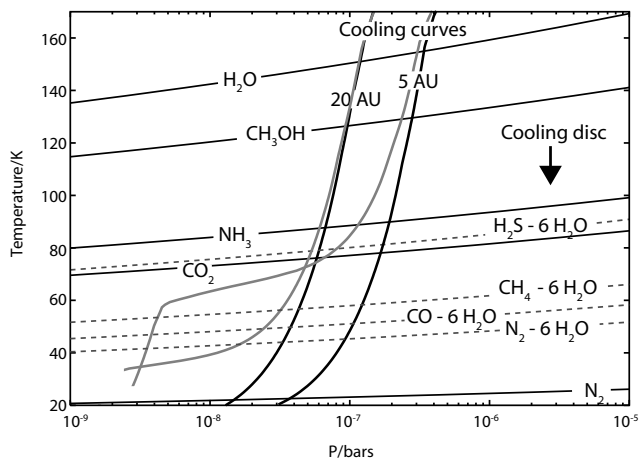}
\caption{Stability curves (horizontal lines) of pure condensates (solid lines) and clathrates (X-6H$_2$O, dashed lines) and cooling curves (vertical lines) of the protoplanetary disc model at two given distances to the star: 5 and 20 AU (the evolution of the disc proceeds from high to low temperatures), for irradiated (grey lines) and non-irradiated (black lines) discs. Species remain in the gas phase above the stability curves. Below, they condense as pure ice or are trapped as clathrates on the surface of grains. Model with CO:CO$_2$ = 1:1.}
\label{cooling_clathrates}
\end{center}
\end{figure}

We note that NH$_3$ does not form stoichiometric hemihydrates and/or monohydrates (NH$_3$ - H$_2$O), but rather a pure condensate whatever the distance to the star, and whatever the chemical composition adopted. This result is in contrast with previous works (Mousis et al. 2010; Marboeuf et al. 2008; Hersant et al. 2004; Lewis 1972; Gautier et al. 2001; Iro et al. 2003) that consider the formation of monohydrate NH$_3$ - H$_2$O in solar nebula, and is due both to the abundance of NH$_3$ adopted in the gas phase of the protoplanetary disc, and to the adoption of a new equilibrium pressure law of condensation of the molecule NH$_3$ (see Table \ref{thermoparam}) described in Fray \& Schmitt (2009). Indeed, the old relation used by other authors overestimates the stability pressure at a given temperature (see Fig.~\ref{NH3}). However, NH$_3$ can still form a stoichiometric hemihydrate and/or monohydrate once the elements have been condensed, but its abundance in the disc is given by the thermodynamic conditions of condensation.

For models with clathrate formation, we assume that H$_2$O molecules at the surface of icy grains can form cages of clathrate structure during the cooling of the stellar nebula. In this case, only H$_2$S, CH$_4$, CO, and N$_2$ molecules are trapped in cages before their condensation at lower temperature (see Fig.~\ref{cooling_clathrates}). CO and N$_2$ can be both trapped (in water ice structure) and condensed at the surface of grains. N$_2$ is trapped in clathrates for the poor CO (CO:CO$_2$ = 1:1) model but does not form clathrates for the rich CO (CO:CO$_2$ = 5:1) model : with the increase in the CO abundance, the amount of H$_2$O molecules free to trap N$_2$ in clathrates decreases because of CO clathrate formation at higher temperature. 
When H$_2$O molecules are no longer available to trap the volatile species $x$ (H$_2$S, CH$_4$, CO, and N$_2$), the clathrate formation process is stopped and the rest of the volatile molecule (mainly CO and/or N$_2$) condenses at lower temperature as pure ices as soon as the temperature of the disc at a given position $r$ is low enough.\\

The particularity of clathrates is to trap species at higher temperature and pressure, and closer to the star than the process of condensation of pure ices.
The above discussion on the trapping/condensation of species on grains assumes that the temperature of the disc reaches very low values. However, the temperature in the inner part of the disc at the time of planetesimal formation (when, in our model, the composition of planetesimal is frozen) can reach values high enough to prevent condensation and/or the formation of clathrates for all or several chemical species. In this case, the condensation and trapping of volatile molecules in the disc is stopped somewhere along the cooling curve at a given distance $r$ to the star. This is particularly true for the irradiated model that always shows temperatures higher than 30 K within the disc radius of 30 AU : in this model, only clathrates allow CO, N$_2$, and CH$_4$ to be trapped in planetesimals (see Sect.~\ref{composition}). In the irradiated models (grey cooling curves), species are condensed/trapped at lower pressure than non-irradiated models (black cooling curves; see Figures \ref{cooling_condensation} and \ref{cooling_clathrates}).

\subsection{Ice line positions}
\begin{figure*}
\begin{center}
\includegraphics[width=15.cm, angle=0]{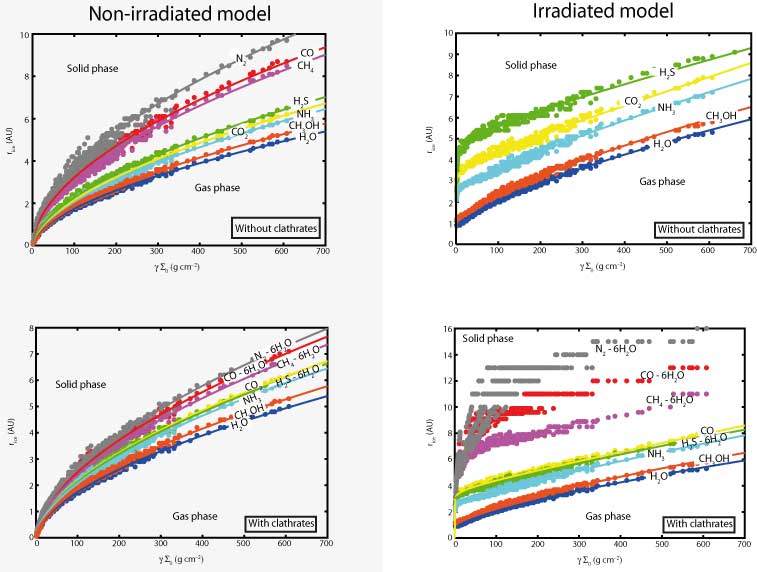}
\caption{Ice and clathrate (lower panels only) line positions (AU) for H$_2$O, CH$_3$OH, NH$_3$, CO$_2$, H$_2$S, CH$_4$, CO, N$_2$, H$_2$S-6H$_2$O, CH$_4$-6H$_2$O, CO-6H$_2$O, and N$_2$-6H$_2$O as a function of $\gamma\Sigma_0$ (g cm$^{-2}$). Dots and lines represent, respectively, data and their interpolation (see Eqs.~\ref{eq:fit_ice_line}, \ref{eq:fit_ice_line_irradiated1}, and \ref{eq:fit_ice_line_irradiated2}, and Tables~\ref{paramfit_noirradiated} and \ref{paramfit_irradiated}).}
\label{fig8}
\end{center}
\end{figure*}

For each volatile molecule, an ice line is formed during the cooling of the stellar nebula (see Fig.\ref{fig8}). The ice line represents the distance from the central star beyond which volatile molecules are in solid phase (ices). It corresponds to the region in the protoplanetary disc where the temperature is low enough to allow condensation of volatile molecules and/or their trapping as clathrates, and their incorporation in planetesimals.  

Figure~\ref{fig8} presents the ice line position (when condensation occurs) and clathrate line position (when clathrate formation occurs) of volatile molecules as a function of the surface density $\Sigma_0$ (g cm$^{-2}$) of the disc at 5.2 AU (varying from 3 g cm$^{-2}$ to about 700 g cm$^{-2}$), and the density slope $\gamma$ (varying from  0.4 to 1.1), for irradiated and non-irradiated discs, both with and without clathrate formation. Between the star and the ice/clathrate line, the volatile molecule is in the gas phase. Beyond the ice/clathrate line, the volatile species is in solid/trapped state (ice). The positions of ice/clathrate line remain valid (with slight changes) whatever the abundance of species in the disc. Whatever the disc model and physical conditions considered during its cooling, the less volatile species H$_2$O, CH$_3$OH, and NH$_3$ are always the first three species to condense, and the species CH$_4$, CO, and N$_2$ the last three to condense/be trapped, as soon as the temperature is low enough. Now we discuss the position of ice lines as a function of physical assumptions considered in the model, especially for irradiated and non-irradiated discs, and clathrate formation of species.

\paragraph{Non-irradiated dics:}
In this model, we do not consider irradiation from the central star as a source of heating. The ice line positions $r_{ice}(x)$ of all the species vary approximately from 0 AU to 5-10 AU as a function of $\gamma ~ \Sigma_0$. The less volatile molecules vary approximately in the ranges 0.06-7 AU for H$_2$O, CH$_3$OH, , NH$_3$, CO$_2$, and H$_2$S. Highly volatile species CH$_4$, CO, and N$_2$ have ice lines position that vary in the ranges 0.9-10 AU. The very low value of ice line position (0.06 AU) corresponds to a disc with a low surface density $\Sigma_0$ of 3 g cm$^{-2}$. As the disc is non-irradiated, its temperature is underestimated in the central parts, especially below 1 AU where the irradiated disc model is physically more appropriate.\\

If the clathrate formation kinetics is favourable in the disc, the formation of clathrates at the surface of grains allows the species H$_2$S, CH$_4$, CO, and N$_2$ to be trapped in water ice before their condensation as pure ice and nearer to the star (up to 1 AU). This means that the abundances of trapped species in planetesimals are higher than the corresponding model with only pure ice condensation (see Sect.~\ref{composition}).

From our results (see Fig.~\ref{fig8}), we find a power-law relation that fits the ice/clathrate line position $r_{ice}(x)$ for each species $x$ in non-irradiated models,
\begin{equation}
r_{ice}(x)=\left(\frac{\gamma~\Sigma_0}{\nu(T_c)}\right)^{\mu(T_c)}   \qquad (AU),
\label{eq:fit_ice_line}
\end{equation}
where $\Sigma_0$ (g.cm$^{-2}$) is the surface density of the disc at 5.2 AU, $\gamma$ the density slope; $\nu(T_c)$ and $\mu(T_c)$, given in Table~\ref{paramfit_noirradiated} for each molecule $x$, are functions of the temperature of condensation/trapping $T_c$ of molecules in the disc. We note that $\nu(T_c)$ decreases with $T_c$ (the increasing volatility of molecules); $\nu(T_c)$ and $\mu(T_c)$ can both be expressed by the equation using $T_c$ at 1 AU (see Tables~\ref{paramfit_noirradiated} and \ref{paramfit2_noirradiated}),
\begin{equation}
X(T_c)=a~ln(T_c) + b~T_c + c,
\label{eq:fit_ice_line2}
\end{equation}
where $X(T_c)$ represents  $\nu(T_c)$ or $\mu(T_c)$, and the parameters $a$, $b$, and $c$ are given in Table~\ref{paramfit2_noirradiated}.\\
The lower the surface density of the disc, the closer to the star is the ice line, simply because the temperature structure of the protoplanetary disc results from viscous heating, which is a growing function of the gas surface density (higher surface density discs are hotter). 
Interestingly enough, such a dependence is not necessarily found for passive discs, where the main source of heating is the irradiation from the central star, which does not depend on the gas surface density. In this case, the location of the ice lines would not depend only on the disc mass (see below and Fig.~\ref{fig8}).

\begin{table}
\centering 
\caption{Parameters allowing the calculation of the ice and clathrate line position $r_{ice}(x)$ (AU) for each volatile molecule $x$ (see Eq.~\ref{eq:fit_ice_line}) in non-irradiated discs.}
\begin{tabular}{l|cc|c}
\hline
\hline
Molecules & $\nu(T_c)$  & $\mu(T_c)$ & $T_c$$^*$ (K)\\ 
\hline
Pure ices &        &      &   \\
  H$_2$O	 & 38.65 $\pm$  0.21  	&  0.582 $\pm$ 0.002	& 160\\
  CH$_3$OH & 35.34 $\pm$ 0.21    	&  0.587 $\pm$ 0.002	& 134\\
  NH$_3$   & 28.93 $\pm$ 0.19	    &	 0.585 $\pm$ 0.002  & 93\\
  CO$_2$   & 26.81 $\pm$ 0.19     &  0.584  $\pm$ 0.002 & 81\\
  H$_2$S   & 24.58 $\pm$  0.19    &  0.582 $\pm$  0.002 & 70 \\
  CH$_4$   & 13.30  $\pm$ 0.18    &  0.555  $\pm$ 0.0023& 33 \\
  CO       & 11.96  $\pm$ 0.18    &  0.55  $\pm$ 0.003  & 29\\ 
  N$_2$    & 8.846 $\pm$ 0.179    &  0.54 $\pm$ 0.0036  & 23\\
  \hline
  Clathrates &        &      &   \\
  H$_2$S - 6 H$_2$O  & 27.21 $\pm$  0.19    &  0.582 $\pm$  0.002 & 84 \\
  CH$_4$ - 6 H$_2$O  & 22.37  $\pm$ 0.19    &  0.579  $\pm$ 0.0022& 61 \\
  CO - 6 H$_2$O      & 20.47  $\pm$ 0.18    &  0.576  $\pm$ 0.0023& 53\\ 
  N$_2$  - 6 H$_2$O  & 18.59 $\pm$ 0.18    &  0.572 $\pm$ 0.0024  & 47\\  
  \hline
 \end{tabular} 
\begin{flushleft}
$^*$Temperature of condensation or clathrate formation of volatile molecules at 1 AU.
\end{flushleft}
\label{paramfit_noirradiated}
\end{table}

\begin{table}
\centering 
\caption{Parameters allowing the calculation of $\nu(T_c)$ and $\mu(T_c)$ (see Eq.~\ref{eq:fit_ice_line2}).}
\begin{tabular}{l|cc}
\hline
\hline  
Parameters & $\nu(T_c)$  																	& $\mu(T_c)$ \\ 
\hline
 a				&	 	12.20 $\pm$ 0.25									  &		  0.062  $\pm$ 0.002 \\
 b        & 4.55 10$^{-2}$ $\pm$ 3.5 10$^{-3}$	&		  -5.74 10$^{-4}$ $\pm$ 3.17 10$^{-5}$ \\
 c        &	-30.53 $\pm$ 0.78                     &		 0.357  $\pm$  0.007    \\
\hline
\end{tabular}
\label{paramfit2_noirradiated}
\end{table}

\begin{figure}[h]
\begin{center}
\includegraphics[width=9.cm, angle=0]{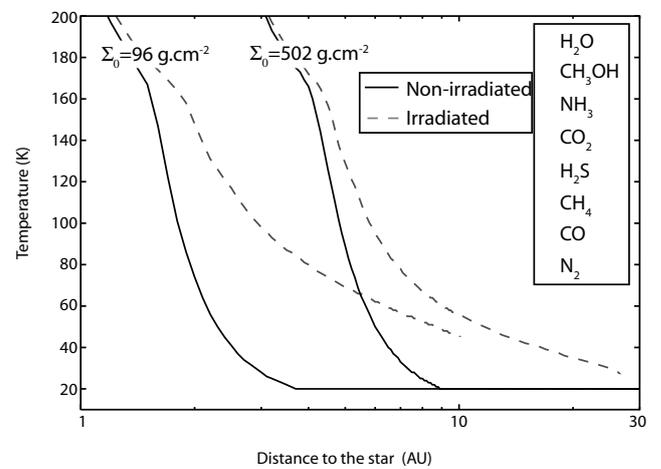}
\caption{Final temperature profile of irradiated (dashed grey lines) and non-irradiated (black lines) discs for two particular simulations ($\Sigma_0=$95.844 g.cm$^{-2}$, $a_{core}=$46, $\gamma=$0.9; and $\Sigma_0=$502.52 g.cm$^{-2}$, $a_{core}=$80, $\gamma=$0.9) as a function of the distance to the star. Coloured dots represent the ice line positions of H$_2$O, CH$_3$OH, NH$_3$, CO$_2$, H$_2$S, CH$_4$, CO, and N$_2$. Models without clathrates.}
\label{fit_temperature_profile_discs}
\end{center}
\end{figure}

\paragraph{Irradiated dics:}
For the irradiated discs, the ice/clathrate line positions are all shifted farther from the star compared to non-irradiated models. The distance to the star is more important for all volatile molecules, especially for highly volatile species.
Thus, if the ice line of H$_2$O is farther of only about 0.5 AU from the star compared to the non-irradiated model, the ice line positions of highly volatile molecules CO, N$_2$, and CH$_4$ show large dispersion as a function of $\gamma ~ \Sigma_0$, varying from 3 AU to more than 16 AU for models with clathrates. 
The irradiated model without clathrate formation never condenses the highly volatile species CH$_4$, CO, and N$_2$ in the area of 30 AU around the star since the temperature never decreases below 30 K.
The formation of clathrates in this case allows the species H$_2$S, CH$_4$, CO, and N$_2$ to be trapped in planetesimals. The difference with the non-irradiated discs comes from the temperature-pressure evolution of the disc (see Figs.~\ref{cooling_condensation} and \ref{cooling_clathrates}). Figure~\ref{fit_temperature_profile_discs} presents the temperature profile of irradiated (dashed grey lines) and non-irradiated (black lines) discs (both without clathrates) for two particular simulations ($\Sigma_0=$95.844 g.cm$^{-2}$, $a_{core}=$46, $\gamma=$0.9; and $\Sigma_0=$502.52 g.cm$^{-2}$, $a_{core}=$80, $\gamma=$0.9) as a function of the distance to the star. The coloured dots represent the temperature of condensation and the positions of ice lines of H$_2$O, CH$_3$OH, NH$_3$, CO$_2$, H$_2$S, CH$_4$, CO, and N$_2$. As expected, the irradiated model presents higher temperatures (and pressure) of the disc for the same distance to the star, and hence gas condensation and/or gas trapping occur far away from the star compared to the non-irradiated discs.

The interpolation of the data of ice/clathrate line positions is somewhat difficult for the irradiated discs. We determine two different relations to fit the ice/clathrate line position $r_{ice}(x)$ of the less volatile species H$_2$O, CH$_3$OH, NH$_3$, CO$_2$, and H$_2$S. The ice/clathrate line positions of volatile molecules H$_2$O, CH$_3$OH, NH$_3$, and H$_2$S - 6 H$_2$O can be fitted by a power law (similar to the one in Eq.~\ref{eq:fit_ice_line}):
\begin{equation}
r_{ice}(x)=\left(\frac{\gamma~\Sigma_0}{a}\right)^{b} + c   \qquad (AU).
\label{eq:fit_ice_line_irradiated1}
\end{equation}
CO$_2$ and H$_2$S species are fitted by a logarithmic law,
\begin{equation}
r_{ice}(x)=a~ln(\gamma~\Sigma_0) + b~(\gamma~\Sigma_0) + c   \qquad (AU),
\label{eq:fit_ice_line_irradiated2}
\end{equation}
where $a$, $b$, and $c$ are given in Table~\ref{paramfit_irradiated} for the two equations.
The clathrate line positions of highly volatile species CH$_4$, CO, and N$_2$ are too spatially dispersed as a function of $\gamma~\Sigma_0$ to find a simple law. This means that the locations of ice lines do not depend only on the disc mass of highly volatile species.

\begin{table}
\centering 
\caption{Parameters allowing the calculation of the ice and clathrate line position $r_{ice}(x)$ (AU) for each volatile molecule $x$ (see Eqs.~\ref{eq:fit_ice_line_irradiated1} and \ref{eq:fit_ice_line_irradiated2}) in irradiated discs.}
\begin{tabular}{l|ccc}
\hline
\hline
Molecules & $a$  & $b$ &   $c$     \\ 
\hline
Pure ices$^*$ &        &      &   \\
\dotfill &  \multicolumn {3} {l}{Using Eq.~(\ref{eq:fit_ice_line_irradiated1})}\\
  H$_2$O	 & 63.74 $\pm$ 0.1  	  &  0.69 $\pm$ 0.005	    & 0.70 $\pm$ 0.01 \\
  CH$_3$OH & 68.41 $\pm$ 1.2    	&  0.73 $\pm$ 0.006	    & 1.05 $\pm$ 0.01 \\
  NH$_3$   & 86.98 $\pm$ 1.87	    &	 0.81 $\pm$ 0.01      & 2.43 $\pm$ 0.02 \\
  \dotfill &  \multicolumn {3} {l}{Using Eq.~(\ref{eq:fit_ice_line_irradiated2})}\\
  CO$_2$   & 0.175 $\pm$ 0.008     &  0.0064  $\pm$ 0.0001    & 2.94 $\pm$ 0.03 \\
  H$_2$S   & 0.33 $\pm$ 0.01     &  0.0051 $\pm$  0.00014     & 3.56 $\pm$ 0.04 \\
  \hline
  Clathrates$^*$ &   \multicolumn {3} {l}{Using Eq.~(\ref{eq:fit_ice_line_irradiated1})}   \\
  H$_2$S - 6 H$_2$O  & 88.19 $\pm$  2.54    &  0.80 $\pm$  0.02 & 3.03 $\pm$ 0.02 \\
  \hline
 \end{tabular} 
\begin{flushleft}
$^*$It is impossible to obtain laws for ice/clathrate line positions of CH$_4$, CO, and N$_2$.\\
\end{flushleft}
\label{paramfit_irradiated}
\end{table}

\subsection{Ice/rock mass ratio}
\begin{figure}[h]
\begin{center}
\includegraphics[width=9.cm]{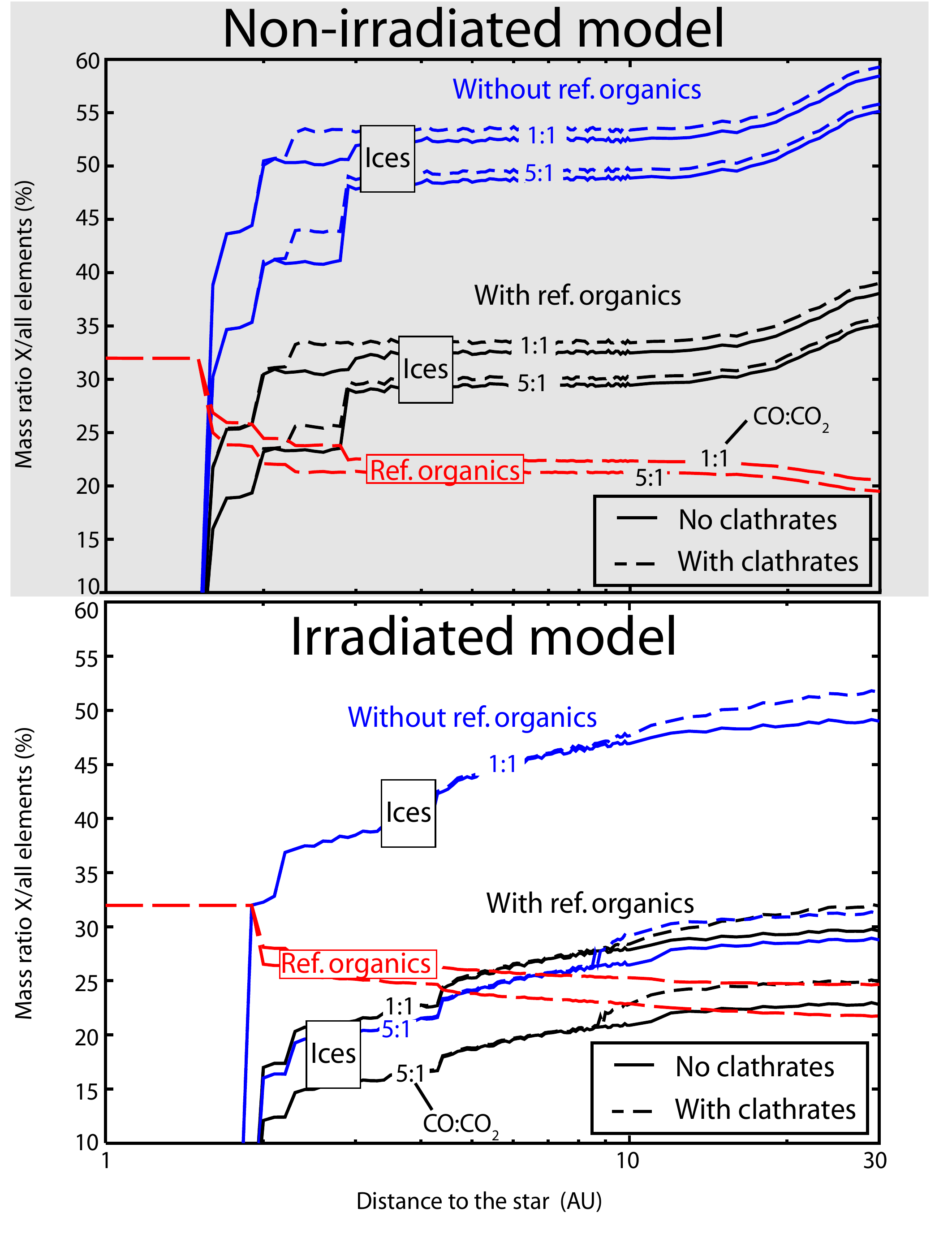}
\caption{Ices and refractory organics mass fractions (relative to the total mass of condensed elements) for one particular simulation ($\Sigma_0=$95.844 g.cm$^{-2}$, $a_{core}=$46, $\gamma=$0.9) as a function of the distance to the star, for non-irradiated (upper panel) and irradiated (lower panel) models, including chemical variation (CO:CO$_2$ molar ratio, refractory organics, clathrate formation).}
\label{ices_all_mass_ratio}
\end{center}
\end{figure}

\begin{figure}[h]
\begin{center}
\includegraphics[width=9.cm]{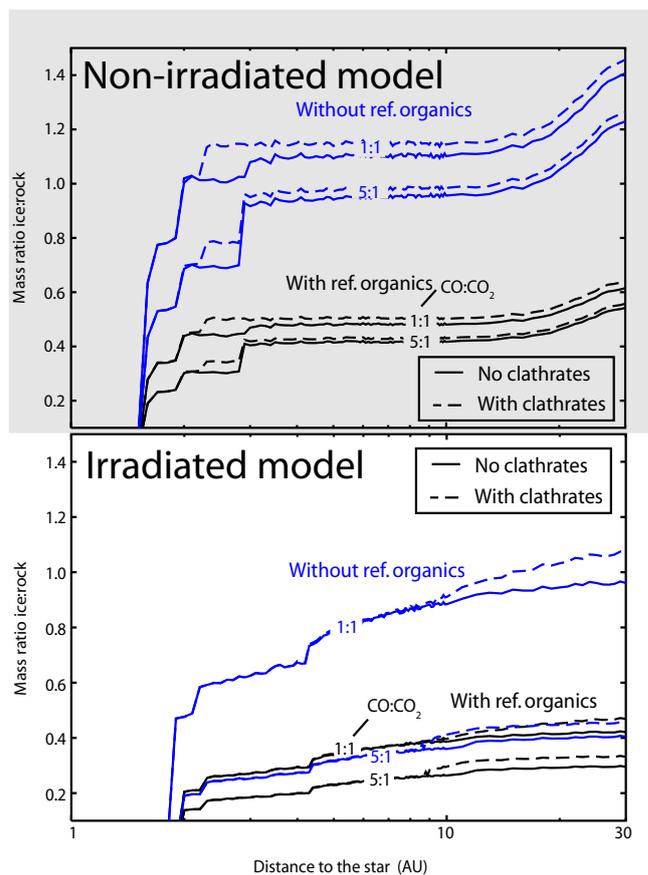}
\caption{Same as Fig.~\ref{ices_all_mass_ratio} but for the ice/rock (rock $=$ minerals + refractory organics) mass ratio in planetesimals.}
\label{ices_rocks_mass_ratio}
\end{center}
\end{figure}

The existence of the ice/clathrate lines in protoplanetary discs induces a chemical differentiation of the discs as a function of the distance to the star. Future planetesimals formed in the discs will be enriched or depleted in volatile molecules following the position of their formation and growth before or after the different ice/clathrate lines. As a result, the ice mass fraction relative to the total mass of condensed elements (resp. ice/rock mass ratio, with rock = minerals + refractory organics) in planetesimals changes as a function of their relative position to the ice/clathrate lines.
Figures \ref{ices_all_mass_ratio} and \ref{ices_rocks_mass_ratio} present, respectively, ice and refractory organics mass fractions relative to the total mass of condensed elements (the sum of ices, minerals, and refractory organics), and the ice/rock mass ratio as a function of the distance to the star, for one particular simulation ($\Sigma_0=$95.844 g.cm$^{-2}$, $a_{core}=$46, $\gamma=$0.9) and for non-irradiated (upper panel) and irradiated (lower panel) models, including chemical variations (CO:CO$_2$ molar ratio, refractory organics, and clathrate formation). 

Whatever the models, we observe a strong  gradient  in  the ice-to-rock mass ratio with the distance from the star, especially near the H$_2$O ice line position. The variations with the distance to the star of both the ice/all elements mass fraction and ice/rock mass ratio (see Figs.~\ref{ices_all_mass_ratio} and \ref{ices_rocks_mass_ratio}) are due to the positions of the ice line of each volatile molecule, H$_2$O being the most abundant species. Whenever a gas condenses (or is trapped as clathrate), the solid surface density increases at the corresponding position in the protoplanetary disc. Considering all physico-chemical assumptions, the amount of ices incorporated in icy planetesimals (resp. the ice/rock mass ratio) beyond the snow line ranges from 10\% to 60\% (see Fig.~\ref{ices_all_mass_ratio}) of their total mass (resp. from 0.1 to 1.5; see Fig.~\ref{ices_rocks_mass_ratio}).
The irradiated models (lower panels in Figs.~\ref{ices_all_mass_ratio} and \ref{ices_rocks_mass_ratio}) always show lower abundances of elements than the non-irradiated models, whatever the initial chemical composition of the disc before its cooling.\\

The ice/all elements mass fraction and ice/rock mass ratio in icy planetesimals vary as a function of several parameters:\\
$\bullet$ First, the mass fraction of elements varies mainly as a function of the assumption on both organic compounds and irradiation in the disc. Without refractory organics, the amount of ices included in planetesimals increases by a factor of about 1.5 compared to models with organics, whatever the irradiation in the disc. For these models, the mass fraction of refractory organics is at a minimum of $\approx$20 wt\% (see Fig.~\ref{ices_all_mass_ratio}) by using Pollack et al. (1994) and Jessberger et al. (1988) repartitions of O, C, and N atoms between minerals, refractory organics, and volatile molecules (see Sect.~\ref{secparam1} and Table~\ref{paramdisc}). 
Following the area of formation of planetesimals in the disc (relative to the different ice line positions), the abundance of elements in icy planetesimals varies in the range 15-60 wt\% (resp. 10-50 wt\%) for ices, 20-30 wt\% (resp. 20-30 wt\%) for refractory organics, and 40-65 wt\% (resp. 45-70 wt\%) for minerals, for non-irradiated (resp. irradiated) models (see Fig.~\ref{ices_all_mass_ratio}). The ice/rock mass ratio in planetesimals varies from 0.2 to 1.5 (resp. 0.1 to 1.1) for non-irradiated (resp. irradiated) models (see Fig.~\ref{ices_rocks_mass_ratio}).

\begin{figure}[h]
\begin{center}
\includegraphics[width=9.cm]{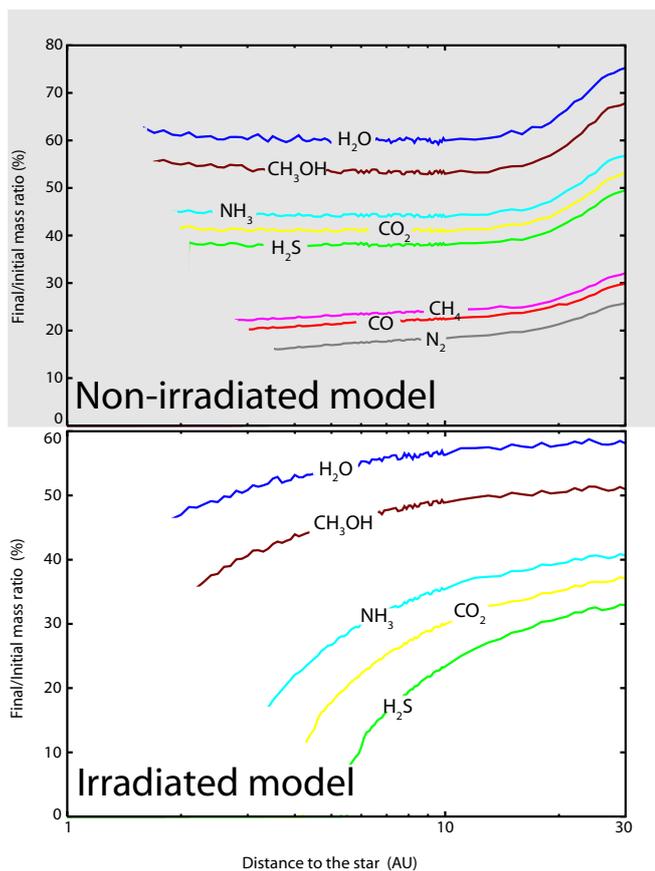}
\caption{Ratio of species $x$ in ices of irradiated (lower panel) and non-irradiated (upper panel) discs (final) relative to the ones in the pristine disc (initial) as a function of the distance to the star. This figure remains valid for the models with CO:CO$_2$ = 1:1 and 5:1. Models without clathrates.}
\label{X_planetesimals_X_ISM}
\end{center}
\end{figure}

\begin{figure}[h]
\begin{center}
\includegraphics[width=9.cm]{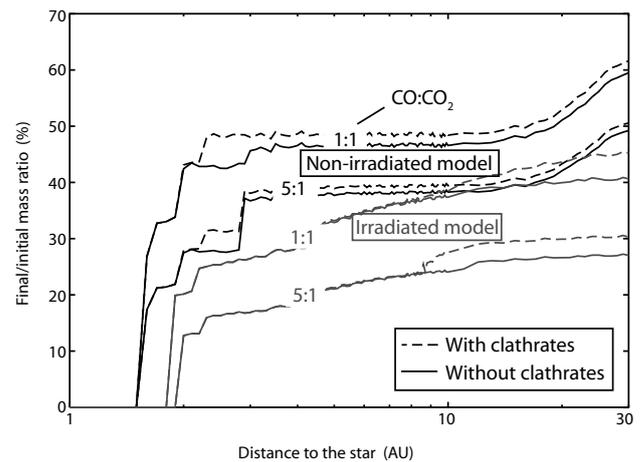}
\caption{Mass ratio of ices in planetesimals (final) relative to the pristine amount of gas and ices in ISM (initial) as a function of the distance to the star, for all models. The black and grey lines represent respectively the non-irradiated and irradiated models, both with (dashed lines) and without (solid lines) clathrates, and models assuming CO:CO$_2$ = 1:1 and 5:1.}
\label{X_planetesimals_ISM}
\end{center}
\end{figure}

$\bullet$ Second, the ice/all elements mass fraction (resp. ice/rock mass ratio) included in icy planetesimals varies as a function of the composition of volatile molecules in the disc (CO:CO$_2$ molar ratio), the formation of clathrates (or not), and the distance to the star, from 1 AU to 30 AU. 
By increasing the fraction of carbon monoxide in the disc (CO:CO$_2$ = 5:1), the total amount of ices in the planetesimals decreases because of the lower temperature of condensation of CO compared to H$_2$O. \\
Figure \ref{X_planetesimals_X_ISM} shows the depletion of volatile molecules in icy planetesimals (final) relative to the ISM (initial) for irradiated (lower panel) and non-irradiated (upper panel) discs as a function of the distance to the star. Compared to the initial amount of ices and gas in the ISM, the abundances of volatile molecules in ices of non-irradiated discs (upper panel) have decreased by $\sim$ 30-60\% for the less volatile species such as H$_2$O, CH$_3$OH, NH$_3$, CO$_2$, and H$_2$S, and by 70-90\% for highly volatile species such as CH$_4$, CO, and N$_2$.
For irradiated discs (lower panel) that represent more physical conditions near the star, these abundances decrease with higher amplitude, and vary even more with the distance to the star (up to several tens of \%). This can be explained by the different profiles and evolutions of the temperature, pressure, and surface density in the disc as shown by figures \ref{cooling_condensation}, \ref{cooling_clathrates}, and \ref{fit_temperature_profile_discs} which are due to different boundary conditions at the surface of the disc, the irradiation or not. For the less volatile species H$_2$O, CH$_3$OH, NH$_3$, CO$_2$, and H$_2$S, the abundances decreases by 40-90\%. For highly volatile molecules CH$_4$, CO, and N$_2$, the abundance can be 0 because the disc is too warm to allow the condensation of these species (but the conditions for their trapping in clathrates are met as soon as the temperature is low enough at a given distance $r$ from the star).
So, highly volatile species such as CO are highly depleted compared to less volatile molecules, especially for irradiated discs. Indeed, by increasing the fraction of CO in the chemical composition of the disc, the mass of ices incorporated in planetesimals decreases for all the models.
This assumption is verified with Fig.~\ref{X_planetesimals_ISM} which shows the depletion of the sum of ices in planetesimals (final) relative to the pristine mass of gas and ices in the ISM (initial) as a function of the distance to the star, for all models. 
The rich CO model of the irradiated disc presents the higher depletion of ices after the cooling of the disc.
Thus, compared to the ISM, the quantity of volatile molecules incorporated in icy planetesimals has decreased by about 40-50\% (resp. 50-70\%) for non-irradiated models assuming CO:CO$_2$ = 1:1 (resp. CO:CO$_2$ = 5:1). The amount decreases by about 60-80\% (resp. 70-90\%) for the irradiated model assuming CO:CO$_2$ = 1:1 (resp. CO:CO$_2$ = 5:1).
The formation of clathrates (dashed line) induces a slight increase in the amount of ices in planetesimals, both for irradiated and non-irradiated models (see Figures \ref{ices_all_mass_ratio} and \ref{ices_rocks_mass_ratio}). With clathrate formation, highly volatile molecules are trapped closer to the star and at temperatures higher than condensation of pure ices. This induces an increase in their abundance in ices, and so an increase in the amount species in planetesimals.

Finally, the ice/all elements mass fraction and ice/rock mass ratio in planetesimals are mainly a function of the amount of refractory organics, the location of the formation in the disc (radiation from the star, position relative to ice line positions) and the abundance of volatile molecules in the disc since it increases/decreases with lower/larger abundances of carbon species. Following these considerations, we find that the ice/rock mass ratio (resp. ice:all elements mass fraction) is mainly equal at 1$\pm$0.5 (resp. 45$\pm$15 wt\%) in icy planetesimals without refractory organics, and 0.35$\pm$0.25 (resp. 30$\pm$10 wt\%) for objects with refractory organics.

\subsection{Chemical composition of ices \label{composition}}

In this section, all molar ratios are relative to the total quantity of ices (the sum of H$_2$O, CO, CO$_2$, CH$_3$OH, CH$_4$, H$_2$S, N$_2$, and NH$_3$ molecules).
The abundance of volatile molecules relative to ices changes with assumptions on the radiation in the disc, the CO:CO$_2$ ratio, the distance to the star with the position of the planetesimals relative to the different ice lines, and the occurrence, or not, of the formation of clathrates. However, the presence of refractory organics does not change the relative abundance of species $x$ in planetesimals. So, the chemical composition given hereafter is valid with and without refractory organics.

\begin{figure}[h]
\begin{center}
\includegraphics[width=9.cm]{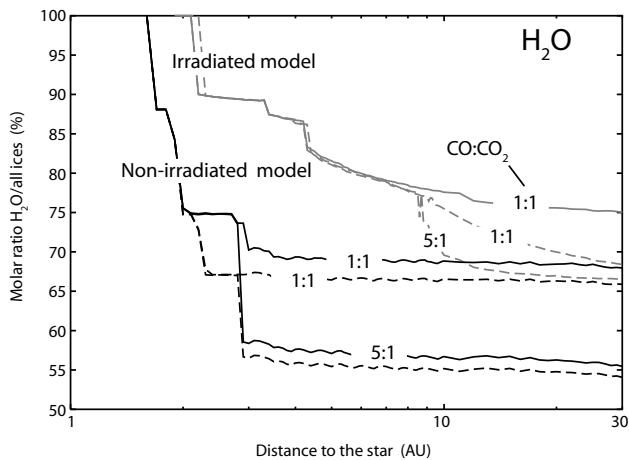}
\caption{Molar ratio of H$_2$O relative to all ices (the sum of all volatile species, including H$_2$O) for one particular simulation ($\Sigma_0=$95.844 g.cm$^{-2}$, $a_{core}=$46, $\gamma=$0.9) as a function of the distance to the star, and for non-irradiated (black lines) and irradiated (grey lines) models, both with (dashed lines) and without (solid lines) clathrates, including chemical variation (CO:CO$_2$ molar ratio).}
\label{fig34}
\end{center}
\end{figure}

\begin{figure}[h]
\begin{center}
\includegraphics[width=9.cm]{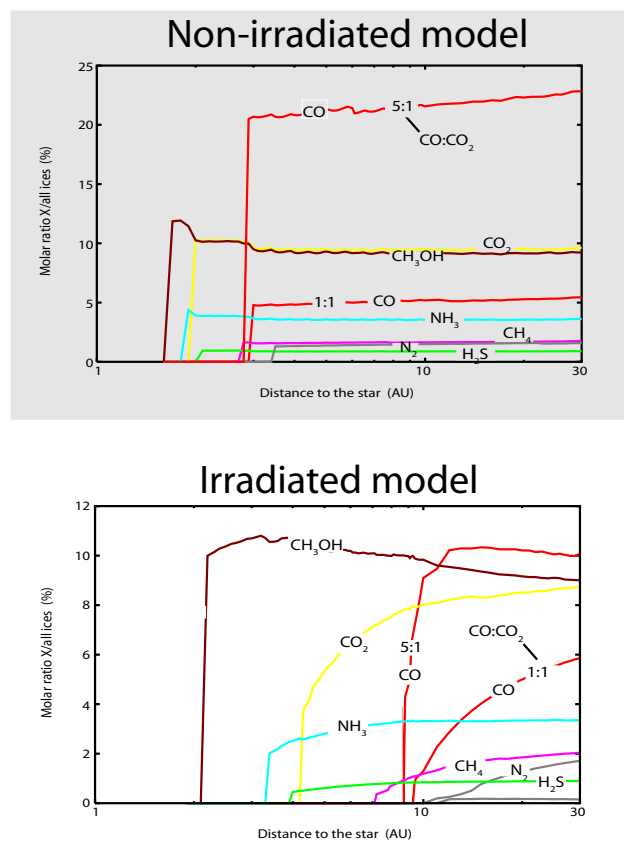}
\caption{Same as Fig.~\ref{fig34} but for species CO, CO$_2$, CH$_4$, CH$_3$OH, H$_2$S, NH$_3$, and N$_2$, and for non-irradiated (upper panel) and irradiated (lower panel) models.}
\label{fig33}
\end{center}
\end{figure}

Figure \ref{fig34} presents the molar ratio of H$_2$O relative to all ices (the sum of all icy species X + H$_2$O) for one particular simulation ($\Sigma_0=$95.844 g.cm$^{-2}$, $a_{core}=$46, $\gamma=$0.9) as a function of the distance to the star, and for non-irradiated (black lines) and irradiated (grey lines) models, both with (dashed lines) and without (solid lines) clathrates, including chemical variation (CO:CO$_2$ molar ratio). Figure \ref{fig33} presents the same molar ratio for species X (X=CO, CO$_2$, CH$_4$, CH$_3$OH, NH$_3$, N$_2$, and H$_2$S) relative to all ices, for non-irradiated (upper panel) and irradiated (lower panel) models.

\noindent Whatever the chemical composition of the disc (CO:CO$_2$ $=$ 1:1 or 5:1), distance to the star (beyond the H$_2$O ice line), and the assumption for clathrate formation (with or without), and radiation in the disc, H$_2$O is the major volatile molecule in icy planetesimals (see Fig.~\ref{fig34} and Table~\ref{paramabundancesplanetesimals}). 
For all models, far away from the star (in the range 10-30 AU), the H$_2$O-to-ices molar ratio ranges at minimum from $\approx$ 55 to $\approx$75\%, and grow up to 100\% near the star (water ice line position) with some variations (due to ice line positions of the other species) with the distance to the star.
The irradiated model always contains larger abundances of H$_2$O relative to other species but with continuous radial variations due to irradiation.
The species H$_2$O and CO are the only molecules whose molar ratio varies significantly with the distance to the star. 
The CO abundance in planetesimals is a function of the CO abundance assumed in the disc (CO:CO$_2$=5:1 for rich CO models, and 1:1 for poor CO models) before the condensation/trapping of species. For rich CO and non-irradiated models, the CO abundance reaches about 22$\pm$2\% in planetesimals. For poor CO models, it decreases at 6$\pm$2\%. Variations are due to the position of planetesimals in the disc. When clathrate formation occurs, the molar fraction of CO increases by only a small percent (1-2\%) in planetesimals. Irradiated model presents lower abundances everywhere in the disc with a strong radial variation.
The X-to-ices molar ratios of the less volatile species CO$_2$, CH$_3$OH, NH$_3$, and H$_2$S change poorly with the distance to the star, except for CO$_2$ in the irradiated model which presents a strong gradient with the distance to the star. In non-irradiated models, their abundances are roughly equal to 10$\pm$1\% for CO$_2$, 10$\pm$2\% for CH$_3$OH, 3.5\% and $\approx$1\% respectively for NH$_3$ and H$_2$S. 
In irradiated models, their abundances are always a few percentage points lower with radial variations and shift compared to non-irradiated models.
The irradiated model presents a gradient of abundances of species in grains with the distance to the star.
Considering all the non-irradiated models, the CH$_4$-to-ices molar ratio is equal at $\approx$2\% in planetesimals and the N$_2$-to-ices molar ratio is 1.5$\pm$0.5\%.
In irradiated models, the abundance of highly volatile species CH$_4$, CO, and N$_2$ are less than 1\% because the temperature never decreases below 30 K. Remember that the irradiated disc is more appropriate to model the hot surface layers and areas near the star. Only the formation of clathrates allows highly volatile species to be trapped in grains.

Finally, the dominant carbon-bearing species are either CO (rich CO model), or CO$_2$ and CH$_3$OH (poor CO model). The dominant nitrogen species is NH$_3$, the NH$_3$:N$_2$ molar ratio varying from 2 to 4.\\

\begin{figure*}
\begin{center}
\includegraphics[width=15.cm]{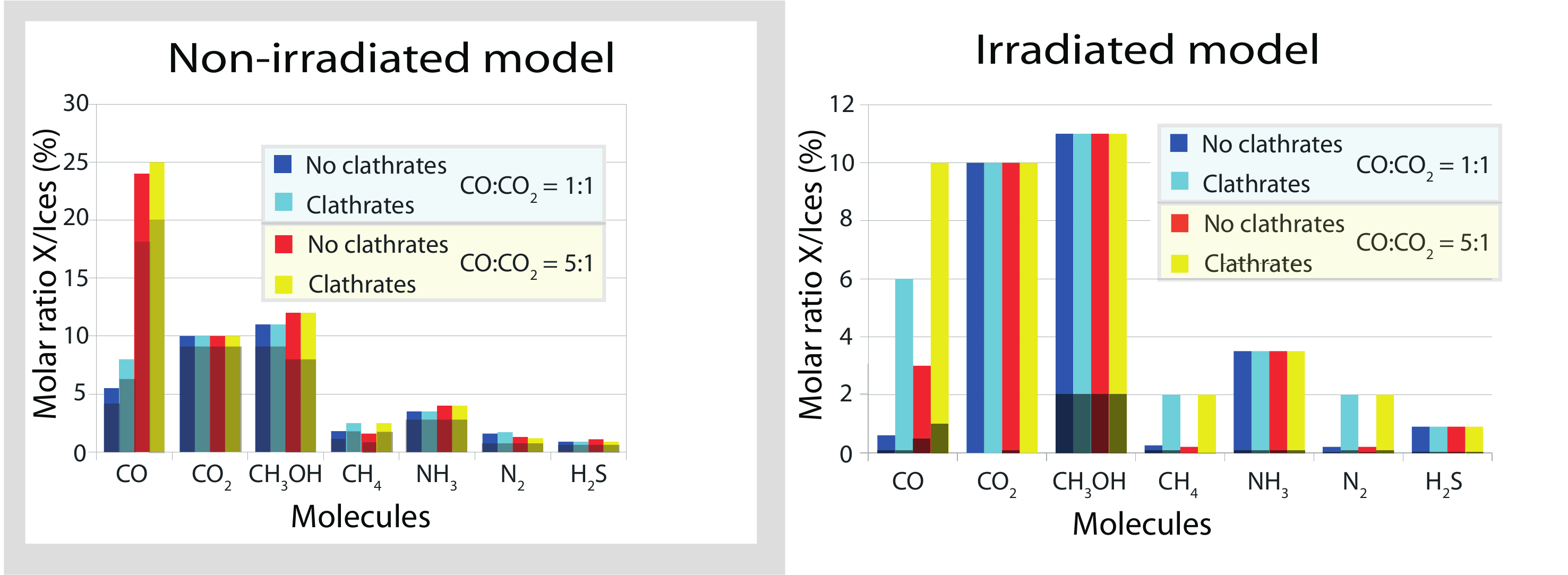}
\caption{Minimum (dark area) and maximum (light area) molar ratios of species $x$ (not H$_2$O) relative to ices (the sum of H$_2$O, CO, CO$_2$, CH$_3$OH, CH$_4$, NH$_3$, N$_2$, and H$_2$S) in icy planetesimals for all models.}
\label{X_ices_planetesimals}
\end{center}
\end{figure*}

\begin{figure}[h]
\begin{center}
\includegraphics[width=9.cm]{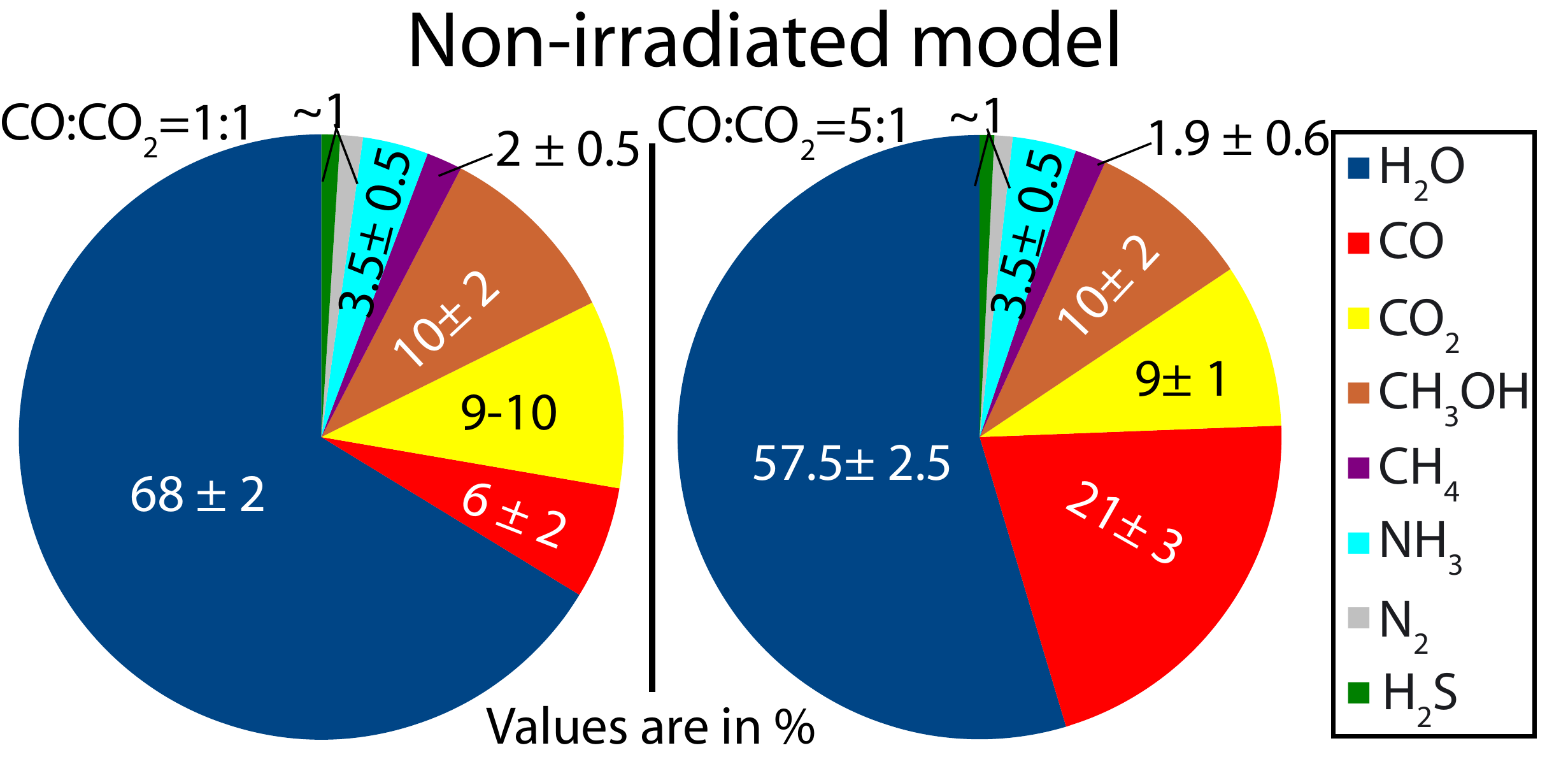}
\caption{Schematic view of the average molar ratio of species $x$ relative to ices in icy planetesimals for all models. The discs are non irradiated.}
\label{X_ices_planetesimals2}
\end{center}
\end{figure}

\begin{figure}[h]
\begin{center}
\includegraphics[width=9.cm]{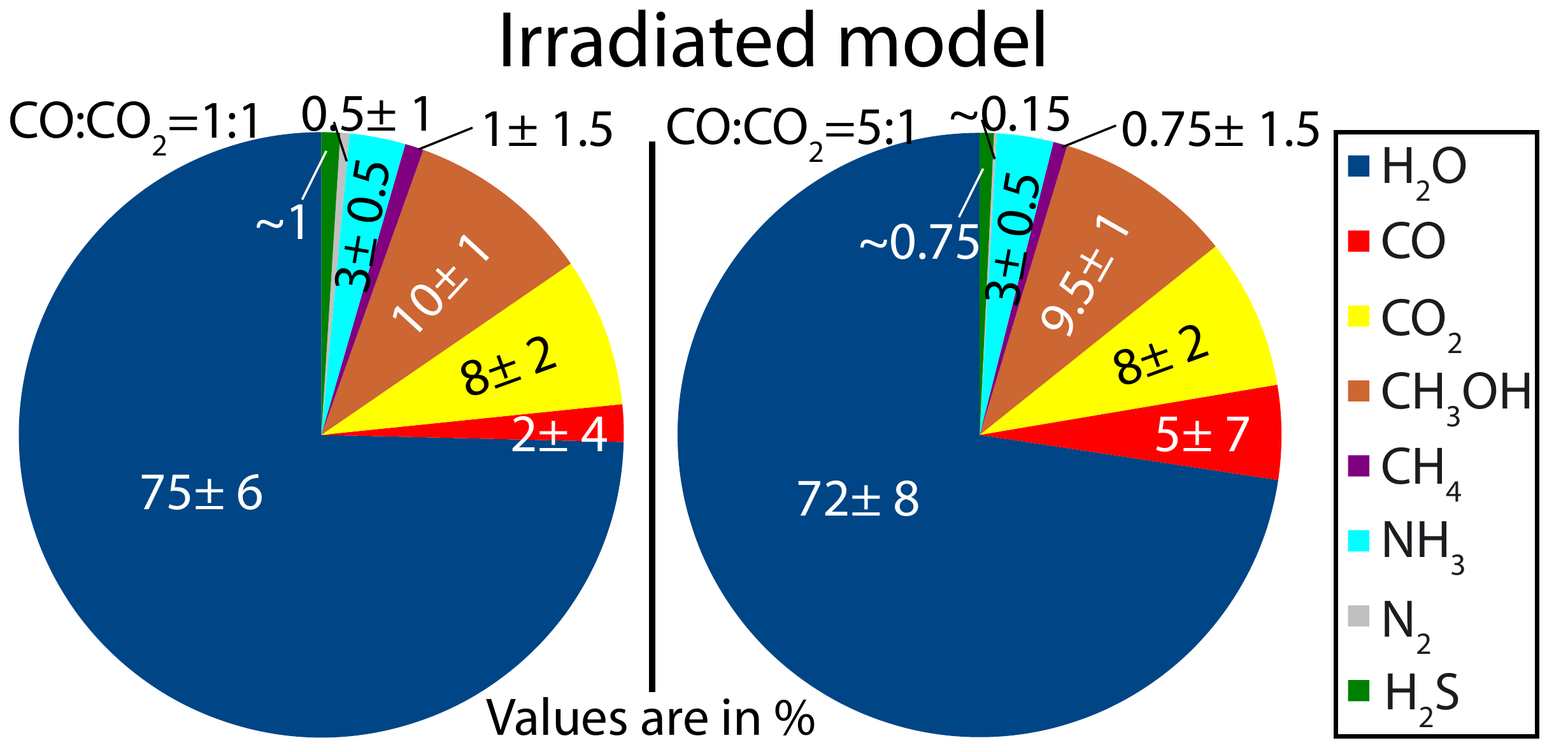}
\caption{Same as in the fig.~\ref{X_ices_planetesimals2} but for irradiated discs.}
\label{X_ices_planetesimals_irradie}
\end{center}
\end{figure}

Figure \ref{X_ices_planetesimals} presents the minimum (dark area) and maximum (light area) molar ratio of species $x$ (except H$_2$O) relative to ices (the sum of H$_2$O, CO, CO$_2$, CH$_3$OH, CH$_4$, NH$_3$, N$_2$, and H$_2$S) in icy planetesimals, for all models. 
Figures \ref{X_ices_planetesimals2} and \ref{X_ices_planetesimals_irradie} present the average molar ratio of all species $x$ relative to ices in icy planetesimals, respectively for all non-irradiated and irradiated models, as a function of the initial chemical composition rich CO and poor CO models. 

Table \ref{paramabundancesplanetesimals} resumes the chemical composition of planetesimals by providing the average molar abundances of species $x$ relative to all ices (and to H$_2$O) for planetesimals beyond the ice line of N$_2$ (the last ice line position in our model) for non-irradiated and irradiated models, and far away from the star where abundance variations are small.
The chemical composition of planetesimals presented in Table~\ref{paramabundancesplanetesimals} and Fig.~\ref{X_ices_planetesimals} remains valid only for bodies located beyond all the ice lines, since they incorporate all species. For bodies located between the H$_2$O ice line and the N$_2$ ice line, the abundance of species relative to ices changes with the distance to the star (position of bodies relative to the different ice lines). Only the abundance relative to H$_2$O (given in Table~\ref{paramabundancesplanetesimals} and Fig.~\ref{comets}) remains valid everywhere beyond the H$_2$O ice line (see Sect.~\ref{comparison_comets} for comparison with comets).

\begin{table*}
\centering 
\caption{Average molar abundance of chemical species in icy planetesimals.}
\begin{tabular}{l|cc|cc||cc|cc||c}
\hline
\hline
												&     \multicolumn {8} {c||}{Planetesimals}  &     Comets\\ 		
\hline
Models				&  \multicolumn {4} {c||}{CO:CO$_2$ = 1:1} & \multicolumn {4} {c||}{CO:CO$_2$ = 5:1} &    \\
\hline
 Molecules (X$_p$)	  &  \multicolumn {2} {c|}{X$_p$/Ices$^*$ (\%)} &  \multicolumn {2} {c||}{X$_p$/H$_2$O (\%)} & \multicolumn {2} {c|}{X$_p$/Ices$^*$ (\%)} &  \multicolumn {2} {c||}{X$_p$/H$_2$O (\%)} & X$_c$/H$_2$O (\%)$^{**}$ \\
\hline
Irradiation			 &	 no     &   yes  &    no    & yes   &   no    & yes   & no & yes&      \\  			
\hline												
H$_2$O	 				 &	68 $\pm$ 2    & 75 $\pm$6  &		100  				  & 100	   	     &	57.5 $\pm$ 2.5  & 72$\pm$8       &	100	          & 100					& -	\\ 
CO							 &	6 $\pm$ 2	    &  2$\pm$4   &	  9$\pm$ 3		 	& 3$\pm$6      &	21 $\pm$ 3 			&	5$\pm$7        & 	37.5$\pm$7.5	&  8$\pm$10	  &0.4-30\\	
CO$_2$	  			 &		9-10        &  8$\pm$2   &	13.5$\pm$ 1.5 	&	11$\pm$3     &  9 $\pm$ 1			  &	8$\pm$2        &	13.5$\pm$1.5	& 11$\pm$3	  & 2-30   \\
CH$_4$	         &	2$\pm$ 0.5    &  1$\pm$1.5 &		3$\pm$ 1			& 1$\pm$2      & 1.9 $\pm$ 0.6		&	0.75$\pm$1.5   &	3$\pm$1	      & 1$\pm$2	    &	0.4-1.6 \\
H$_2$S	         &	$\approx$ 0.9 & 1$\pm$0.25 &    1.2$\pm$ 0.2	& 1$\pm$0.5    & 0.8 $\pm$ 0.1		&	0.75$\pm$0.25  &	1.2$\pm$0.2		& 1$\pm$0.5   &	0.12-1.4\\
N$_2$	           &	1.2 $\pm$ 0.4 & 0.5$\pm$1  & 		1.9$\pm$ 0.7	& 0.7$\pm$1.5  &  1 $\pm$ 0.3  	  &	0.15$\pm$0.035 & 1.9$\pm$0.7    & 0.2$\pm$0.05&	no data \\
NH$_3$	         & 3.5 $\pm$ 0.5  &  3$\pm$0.5 &		$\approx$ 5		& 4.5$\pm$0.75 &  3.5 $\pm$ 0.5   &	3$\pm$0.5      & $\approx$5 	  & 4.5$\pm$1   &  0.2-1.4 \\
CH$_3$OH         &	10 $\pm$ 2	  &  10$\pm$1  &		$\approx$ 13  & 13$\pm$0.3   & 10 $\pm$ 2 		  &	9.5$\pm$1      & $\approx$13 	  & 13$\pm$0.5  &	0.2-7 \\
\hline
\hline
\end{tabular}
\begin{flushleft}
$^*$Ices= H$_2$O + CO + CO$_2$ + CH$_4$ +  H$_2$S +  N$_2$ +  NH$_3$ +  CH$_3$OH \\
$^{**}$data from Mumma \& Charnley (2011) \\
\end{flushleft}
\label{paramabundancesplanetesimals}
\end{table*}

\begin{figure}[h]
\begin{center}
\includegraphics[width=9.cm]{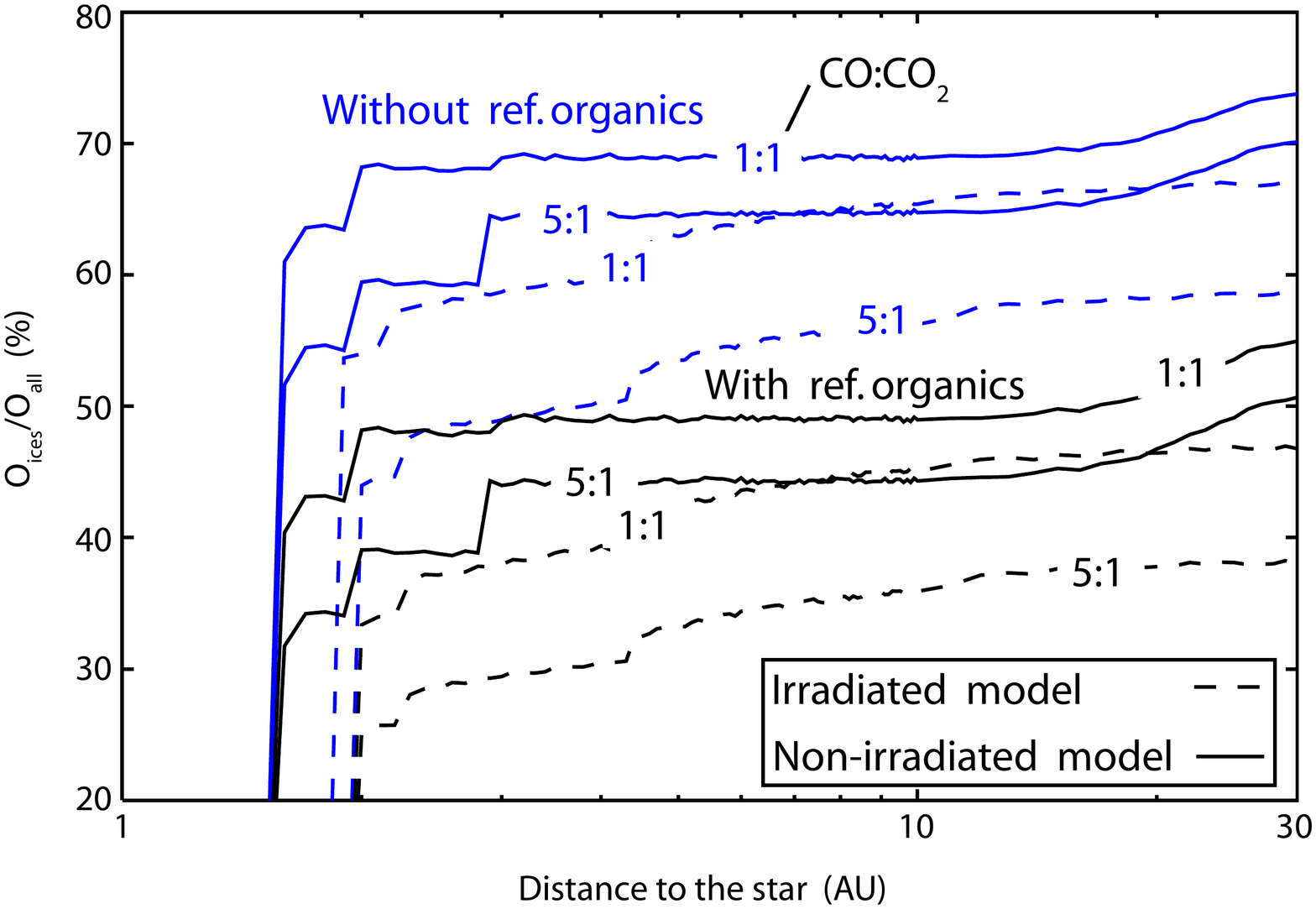}
\caption{Ratio of atoms O$^{ices}$ in ices relative to atoms O$^{all}$ in all solid components of grains, for one particular simulation ($\Sigma_0=$95.844 g.cm$^{-2}$, $a_{core}=$46, $\gamma=$0.9), as a function of the distance to the star, and for non-irradiated (full lines) and irradiated (dashed lines) models, including chemical variation (CO:CO$_2$ molar ratio), and with (blue lines) and without (black lines) refractory organics.}
\label{fig35a}
\end{center}
\end{figure}

\paragraph{Abundances of atoms O and C, and C:O ratio in ices}
We provide the abundances of O and C atoms in ices incorporated in planetesimals relative respectively to total O and C atoms (including minerals, refractory organics and volatile molecules, see Table~\ref{paramabundancesatoms}).
Figure \ref{fig35a} shows the ratio of O$^{ices}$ atoms in ices relative to the O$^{all}$ atoms in all solid components of grains, for one particular simulation ($\Sigma_0=$95.844 g.cm$^{-2}$, $a_{core}=$46, $\gamma=$0.9), as a function of the distance to the star, and for non-irradiated (full lines) and irradiated (dashed lines) models, including chemical variation (CO:CO$_2$ molar ratio), and with (blue lines) and without (black lines) refractory organics.
Planetesimals without refractory organics show higher ratios than models with. The oxygen in ices can represent up to $\approx$ 75\% of all the oxygen in the planetesimal far away from the star. Near the star, in irradiated area of discs, it decreases to $\approx$ 50\%.
Models with refractory organics show ratios from 25\% near the star (irradiated area of discs) to 55\% far away from the star (non-irradiated are of discs). 
Taking into account all distances to the star and physico-chemical conditions, ices incorporate approximately 55$\pm$20\% of total O atoms in planetesimals (including minerals, organics and volatile molecules).

\begin{figure}[h]
\begin{center}
\includegraphics[width=9.cm]{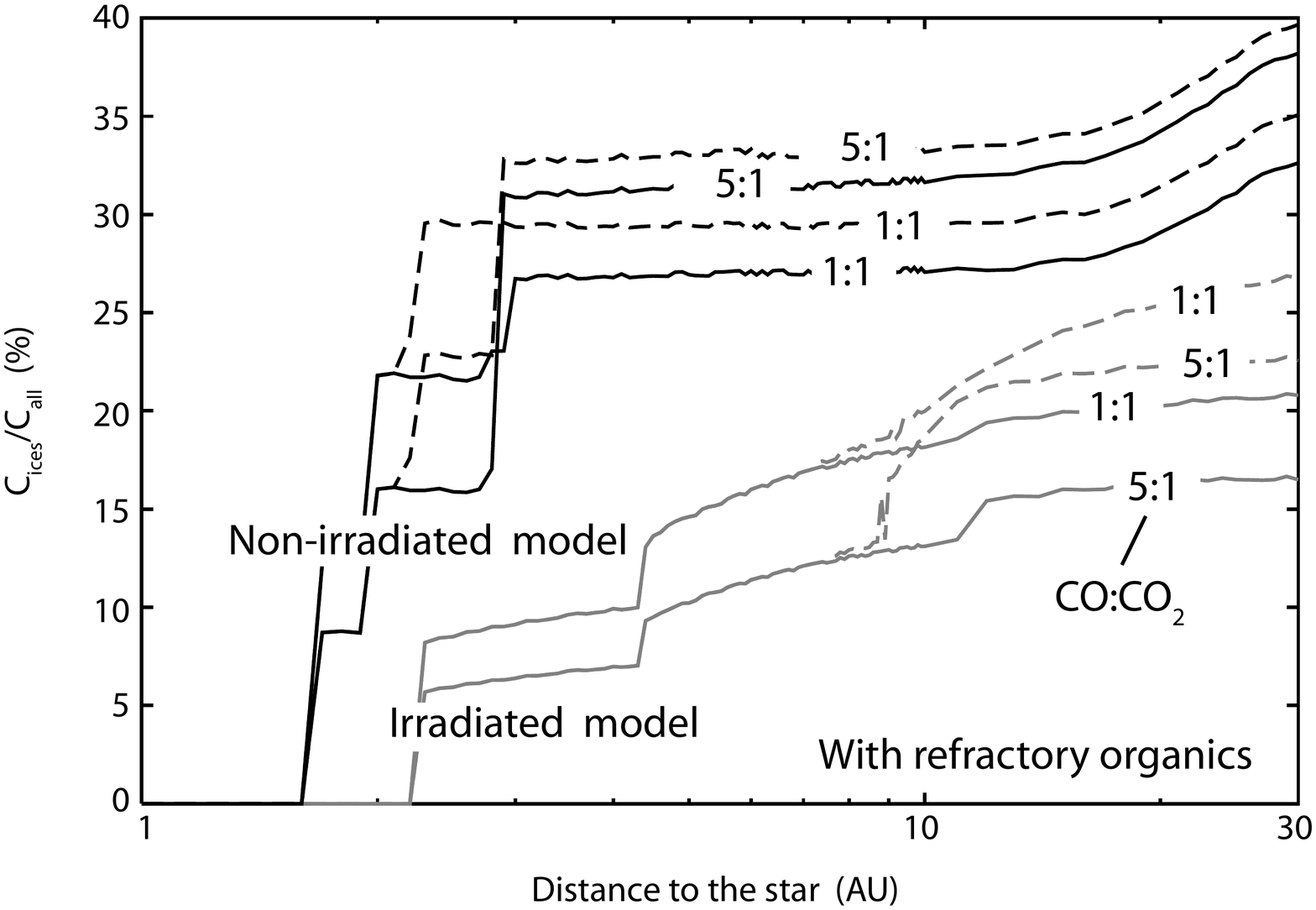}
\caption{Same as Fig.~\ref{fig35a}  but for atoms C$^{ices}$ in ices relative to atoms C$^{all}$ in all solid components of grains.}
\label{fig35b}
\end{center}
\end{figure}

Figure \ref{fig35b} presents the ratio of atoms C$^{ices}$ in ices relative to atoms C$^{all}$ in all solid components of grains, for one particular simulation ($\Sigma_0=$95.844 g.cm$^{-2}$, $a_{core}=$46, $\gamma=$0.9), as a function of the distance to the star, and for non-irradiated (full lines) and irradiated (dashed lines) models, including chemical variation (CO:CO$_2$ molar ratio), and only with refractory organics. When refractory organics are not present in planetesimals, all the carbon is in the ices. In this case, the fraction is 1. With refractory organics, the ratio is approximately 30$\pm$10 \% for non-irradiated models. It decreases to about 15$\pm$10\% for irradiated models.

\begin{figure}[h]
\begin{center}
\includegraphics[width=9.cm]{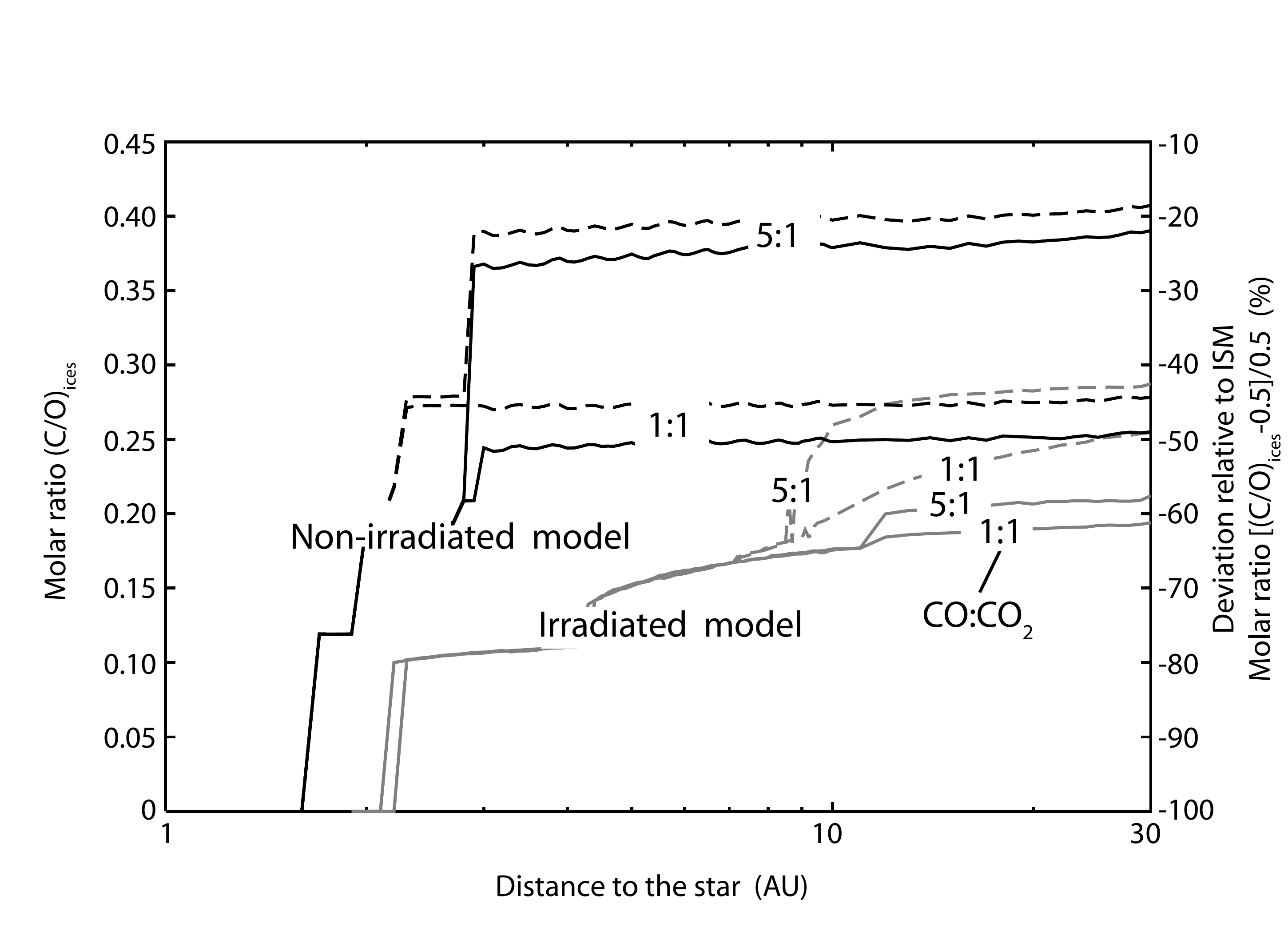}
\caption{Molar ratio C:O in ices (left scale) and its deviation relative to the solar abundance (right scale), for one particular simulation ($\Sigma_0=$95.844 g.cm$^{-2}$, $a_{core}=$46, $\gamma=$0.9), as a function of the distance to the star, and for non-irradiated (black lines) and irradiated (grey lines) models, including chemical variation (CO:CO$_2$ molar ratio), and with (dashed lines) and without (solid lines) clathrates.}
\label{fig36}
\end{center}
\end{figure}

Figure \ref{fig36} shows the molar ratio C:O in ices (left scale) and its deviation relative to the solar abundance (right scale), for one particular simulation ($\Sigma_0=$95.844 g.cm$^{-2}$, $a_{core}=$46, $\gamma=$0.9), as a function of the distance to the star, and for non-irradiated and irradiated models, including chemical variation (CO:CO$_2$ molar ratio), and with and without clathrates. Remember that the C:O molar ratio in the early stellar system (solar system composition) is about 0.5 (see Table~\ref{paramdiscatoms}). Planetesimals formed in non-irradiated discs present the higher C:O ratios with values up to 0.4. In irradiated discs, the ratio decreases to 0.1. Considering all values, the C:O molar ratio in ices is 0.25$\pm$0.15 with deviations relative to the ISM value (solar value in this study) between -20\% and -80\%. Variations include the variation of CO abundance in models and the assumption on the clathrate formation. Values are valid with and without refractory organics. The C:O ratio in ices of planetesimals and future comets should be always in depletion relative to the ISM value.

\begin{figure}[h]
\begin{center}
\includegraphics[width=9.cm]{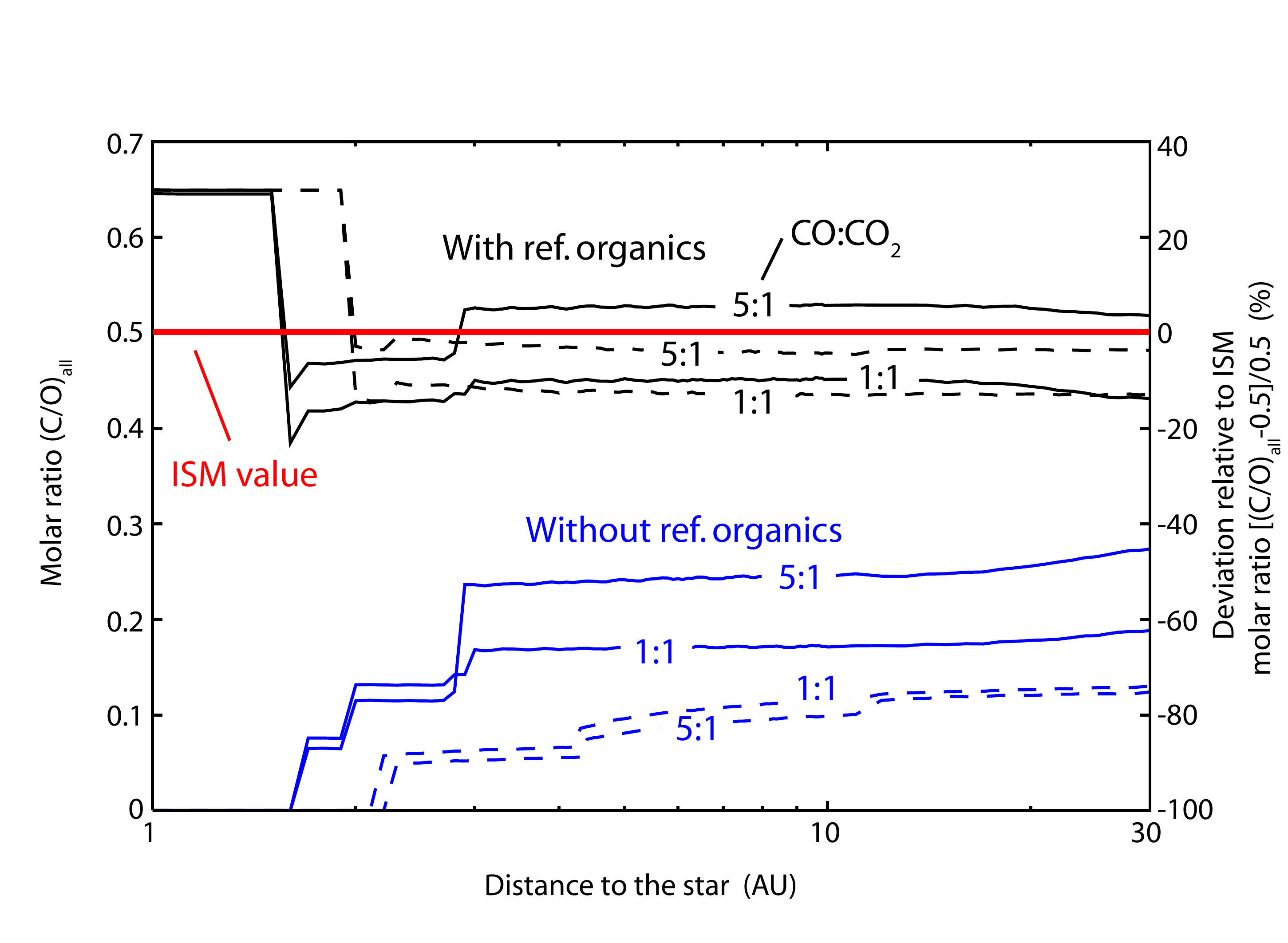}
\caption{Same as Fig.~\ref{fig36} but for molar ratio C:O in all solid components of grains. Solid and dashed lines represent respectively the non-irradiated and irradiated models.}
\label{fig37}
\end{center}
\end{figure}

Figure \ref{fig37} shows the molar ratio C:O in all solid components of grains (left scale) and its deviation relative to the solar abundance (right scale), for one particular simulation ($\Sigma_0=$95.844 g.cm$^{-2}$, $a_{core}=$46, $\gamma=$0.9), as a function of the distance to the star, and for non-irradiated and irradiated models, including chemical variation (CO:CO$_2$ molar ratio), and with and without clathrates. Planetesimals with refractory organics show C:O molar ratio higher than 0.4 with less than $\pm$20\% of deviation relative to the ISM value (solar value in this study). Planetesimals without refractory organics show molar ratio inferior to 0.3 with deviation higher than 40\% for all models. If refractory organics formed before the condensation of water ice are negligible in the cooling disc, planetesimals and future planets (see paper 3) present high C:O depletion relative to the initial ISM value (in our study, the solar value).
\begin{table}
\centering 
\caption{Molar abundance of atoms $x$ in ices relative to total atoms $x$ in planetesimals (X$^{ices}_p$/X$^{all}_p$) for models with and without refractory organics.}
\begin{tabular}{l|cc}
\hline
\hline
Models	&     with  & without \\
\hline
X$^{ices}_p$					    \\	
\hline														
O$^{ices}$/O$^{all}$ 			(\%)						 		&		40$\pm$15   		&	65$\pm$10  	\\ 
C$^{ices}$/C$^{all}$ 			(\%)						     &		22 $\pm$18	  &			1					\\	
\hline
 C/O					 (ices)        &  \multicolumn {2} {c}{0.25$\pm$0.15}   \\
  C/O					 (all solid components)  & 0.47$\pm$0.07 & 0.2$\pm$0.1   \\
\hline
\end{tabular}
\label{paramabundancesatoms}
\end{table}

\section{Comparison to comets \label{comparison_comets}}
\begin{figure}[h]
\begin{center}
\includegraphics[width=8.cm]{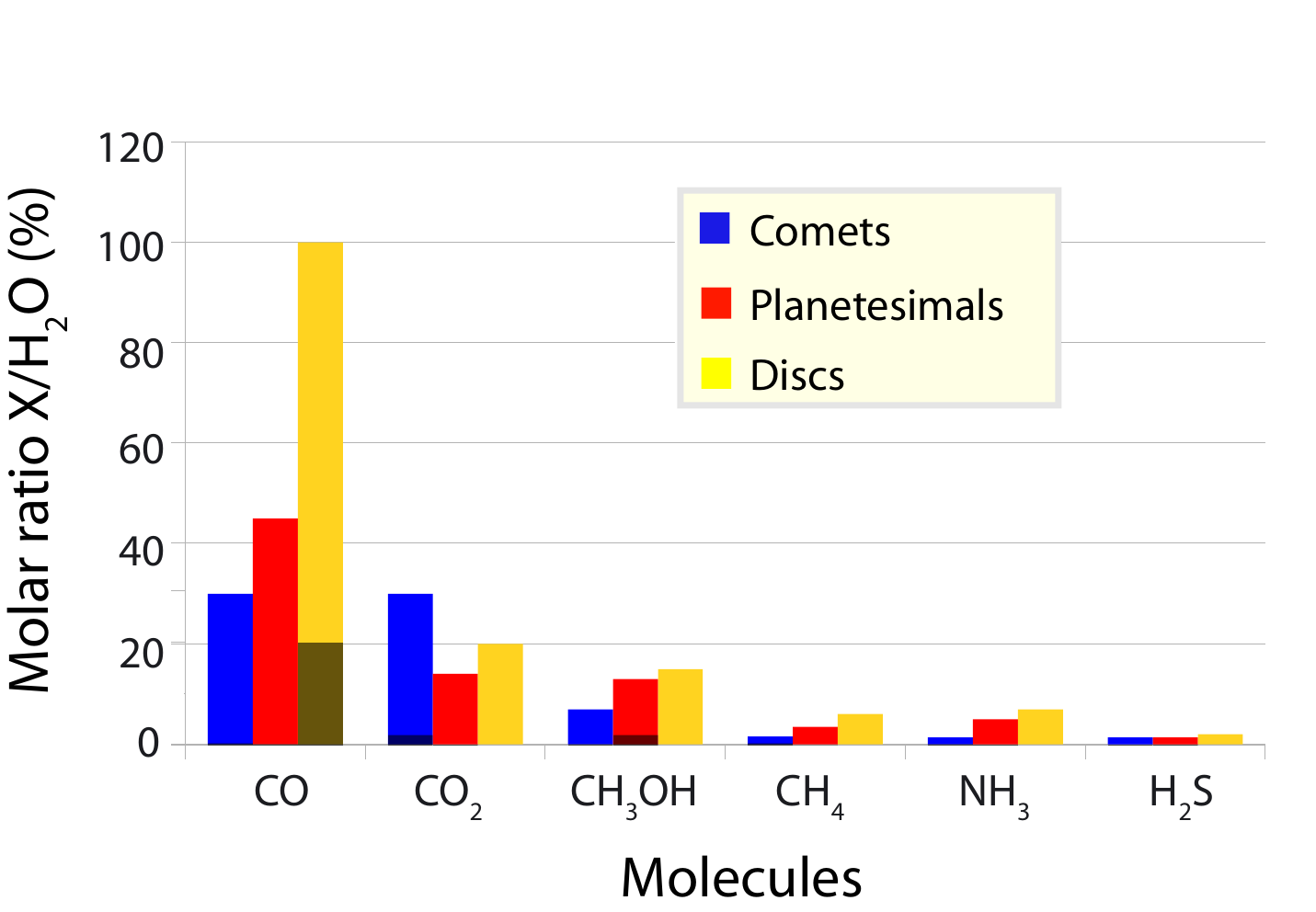}
\caption{Minimum (dark area) and maximum (light area) molar ratios of species $x$ relative to H$_2$O from the gas phase of the disc (yellow area), icy planetesimals (red area), and observations of comets (blue area). Results take into account all the irradiated and non-irradiated models.}
\label{comets}
\end{center}
\end{figure}

For comparison with the values observed in the coma of comets, Fig.~\ref{comets} shows the maximum (light area) and minimum (dark area) molar ratios of species CO, CO$_2$, CH$_3$OH, CH$_4$, NH$_3$, and H$_2$S relative to H$_2$O for gas phase of the disc model (yellow area), planetesimals (red area), and comets of solar system (blue area). Results take into account all the irradiated and non-irradiated models. Data of comets come from Mumma \& Charnley (2011).

The abundances (in mol) of species (relative to H$_2$O) incorporated in icy planetesimals, for both irradiated and non-irradiated models, frame the values observed in the coma of comets (see Table~\ref{paramabundancesplanetesimals} and Fig.~\ref{comets}; Bockel\'ee-Morvan et al. 2004; Mumma \& Charnley 2011), except for CO$_2$ for which abundances calculated from our models show lower values. This could be explained by higher abundances of CO$_2$ in the gas phase of the ISM or chemical reactions producing CO$_2$ in the disc but not taken into account in this study. It should be kept in mind that we consider only average abundances of species in the ISM, except for CO for which rich and poor models were tested. \\
Table~\ref{paramabundancesplanetesimals} provides molar ratio of the different species $x$ in planetesimals relative to H$_2$O, for all non-irradiated models.
The X:H$_2$O molar ratio is $\approx$13.5$\pm$1.5\% for CO$_2$, and 1.2$\pm$0.2\% for H$_2$S. Some variations of the molar ratio CO:H$_2$O are observed as a function of the initial chemical composition of the stellar nebula and chemical processes considered in the discs (the CO:CO$_2$ ratio, and/or clathrate formation process). Taking into account all the variations with the distance to the star and the formation of clathrates, the CO:H$_2$O molar ratio is 9$\pm$2\% (resp. 37$\pm$7\%) assuming CO:CO$_2$ = 1:1 (resp. CO:CO$_2$ = 5:1). If the rich CO models seem to present values of CO that are too high in planetesimals in regards to comets of the solar system, the poor CO models present abundances in agreement with observations of comets.
Compared to comets, the abundances of CH$_4$, NH$_3$, and CH$_3$OH in icy planetesimals are larger by a factor of up to $\approx$ 2 for CH$_4$, $\approx$ 3.5 for NH$_3$, and 2 for CH$_3$OH. The X:H$_2$O molar ratio is approximately equal to 13\% for CH$_3$OH, 5\% for NH$_3$, and 3$\pm$1\% for CH$_4$. Variations take into account the chemical variation (ratio CO:CO$_2$) in the gas phase of the disc and the assumption on the formation of clathrates. For CH$_4$, only the case without clathrate is in good agrement with observations of comets.\\

It should be remembered in this context that the abundances relative to H$_2$O in comets have to be taken with caution. Indeed, outgassing of volatile molecules from comets is a function of their sublimation temperature and abundance in the nucleus, and also depends on the chemical processes of trapping or release, which themselves are functions of the structure of water ice (amorphous, crystalline or clathrate; see Marboeuf et al. 2012). As a result, abundances observed in coma can vary as a function of the distance to the sun and initial abundances in the nucleus. 
The differences between our model and the observations in comets can result from the adoption of an excessive molecular abundance of these elements in the stellar nebula compared to that in protosolar nebula. Finally, other processes like chemical conversion of these molecules, could also alter the expected abundance of volatile molecules. 

The ice/rock mass ratio of icy planetesimals ($\approx$1$\pm$0.5) without refractory organics frames the ice-to-dust mass ratio (a proxy of the ice/rock mass ratio) indicated for the comet 1P/Halley (mass ratio ice/rock $\approx$ 1) by the Giotto-DIDSY measurements (McDonnell et al. 1987). It is also in agreement with the values prescribed by the Greenberg (1982) interstellar dust model (Tancredi et al. 1994) and given by Lodders (2003) for solar system and photospheric compositions. Models with refractory organics seem to be in disagreement with observations as the ice/rock mass ratio is too low (0.35$\pm$0.25).
However, the ice/rock mass ratio in comets has to be taken with caution. The mass of dust ejected from comets is mainly function of the flow of gas escaping from their surface (i.e. the distance to the sun) and size of grains. A mass fractionation can occur with larger grains which stay at the surface of comets, increasing artificially the ice/rock mass ratio in comets. So, models with refractory organics could match the ratio in comets. Note also that a lower amount of organics in grains could match more closely the ice/rock ratio in comets. Remember that the amount of organics in our model is given by the amount of carbon taken - which varies from 45\% to 75\% (60\% assumed in this work) - from the total carbon reservoir in molecular clouds (see Sect.~3.1 - Pollack et al. 1994; Sekine et al. 2005). By decreasing the amount of refractory organics, the amount of ices increases.

Interestingly, the ice/rock mass ratio derived in our models is 1.5 to 3 times lower than the value generally assumed in planet formation models (Hayashi 1981). The model of Hayashi (1981) is derived from solar atomic abundances of Cameron (1970) that are higher by 50 - 90 \% for atoms C, O, and N compared to Lodders's (2003) abundances. They made some assumptions on the volatile species, taking into account only H$_2$O, CH$_4$, and NH$_3$, all condensing together at 170 K. Moreover, they do not take into account the possibility of forming refractory organic compounds. Finally, all assumptions and parameters adopted by Hayashi (1981) lead to an increase in about 60\% in the mass of ices incorporated in planetesimals compared to our study.

\section{Summary and discussion \label{conclusion}}

In this work, we have computed the composition of ices incorporated in planetesimals using 16 different physico-chemical conditions and different initial masses of discs. We used models of disc accretion and ice formation to calculate the abundance of ices incorporated in planetesimals for different initial physico-chemical conditions varying the abundance of CO in the disc, the amount of refractory organics, the formation of clathrates, the star radiation, and the mass of the disc.
The main assumptions that allowed the calculation of the chemical composition of icy planetesimals are given below.

\begin{enumerate}
	\item Atoms:
\begin{itemize}
  \item O, C, N, and S have solar abundances and are shared among minerals, refractory organics, and volatile molecules 
  \item Two extreme distributions of atoms are studied: the first with refractory organic compounds, the second without. 
\end{itemize}
\item Chemical composition of the disc:
\begin{itemize}
	\item The chemical species are represented only by H$_2$O, CO, CO$_2$, CH$_3$OH, CH$_4$, N$_2$, NH$_3$, and H$_2$S which are the most abundant volatile species observed in ISM and comets of the solar system. 	
	\item Chemical composition and abundances of volatile molecules relative to H$_2$O adopted in the gas phase of the disc are average values provided by observations in ISM.
	\item The molar ratio CO:CO$_2$ in volatile molecules of the gas phase has been varied in the range 1:1-5:1, in agreement with observations of ISM.
	\item All ices and gas species initially in the ISM are assumed to be initially in the gas phase of the disc, implying that the temperature is high enough to sublimate all ices from 0 AU to 30 AU, the area of formation of comets and planets.
	\item The model assumes no chemistry in the gas phase of the disc.	
	\item Ices are formed from pure condensates and/or clathrates produced during the cooling of the disc.
	 \item We use two extreme models of ice formation: 
	 \begin{enumerate}
	 \item the first assumes no clathrate formation (all volatile molecules condense only)
	 \item the second assumes clathrate formation during the cooling of the disc (volatile molecules condense and/or are trapped).
	 \end{enumerate}
	\item The thermodynamic conditions used to compute the cooling of the disc and the formation of ices are the ones derived from an $\alpha$ disc considering both models with and without irradiation.
	\item The grains and planetesimals do not migrate radially through the disc; their chemical composition is only a function of the distance to the star and the mass of the disc.
\end{itemize}
\end{enumerate}

\paragraph{Ice line position and chemical composition of icy planetesimals}

We have shown that ices included in planetesimals are mainly composed of H$_2$O and carbon-bearing species. The main results are given below.
\begin{enumerate}
\item The ice line position of most of the volatile species follows a power law as a function of the initial surface density $\Sigma_0$ of the disc at 5.2 AU.
\item Ice/rock and Ice/all elements ratios:
	\begin{itemize}
		\item The ice/rock mass ratio and abundance of ices incorporated in planetesimals vary as a function of the distance to the star.
		\item The ice/rock mass ratio (resp. ice/all elements mass fraction) is 1$\pm$0.5 (resp. 45$\pm$15 wt\%) in icy planetesimals without refractory organics, in good agreement with the ratio of the solar comet 1P/Halley, and 0.35$\pm$0.25 (resp. 30$\pm$10 wt\%) for objects with refractory organics. We discuss below the implications of the values of these ratios.
	\end{itemize}
\item Chemical composition of planetesimals:
\begin{itemize}
		\item Abundances of volatile molecules are depleted compared to pristine amounts of gas and ices of ISM as a function of their temperature of condensation, distance to the star, and initial abundance in the gas phase. They are depleted by $\sim$ 30-60\% for the less volatile species such as H$_2$O, CH$_3$OH, NH$_3$, CO$_2$, and H$_2$S, and by 70-90\% for highly volatile species such as CH$_4$, CO, and N$_2$.
		\item The molar abundance of chemical species incorporated in icy planetesimals are in good agreement with values observed in the coma of comets, excepted for CO rich environments (CO:CO$_2$=5:1):\\
		\begin{enumerate}
			\item The dominant chemical species remains H$_2$O ranging from 65$\pm$10\% to 100\% (in mol relative to all ices).	
			\item The three dominant carbon-bearing species are CO, CO$_2$, and CH$_3$OH.		
			\item The dominant nitrogen-bearing species is NH$_3$ which is $\approx$ 2-4 $\times$ N$_2$.
			 \item The C:O molar ratio in ices is $\approx$ 0.25$\pm$0.15 considering all values, about 50$\mp$30\% of the ISM value.
			 \item The C:O molar ratio in all solid components of planetesimals is $\approx$ 0.47$\pm$0.07 in icy planetesimals considering the refractory organics compounds, and 0.2$\pm$0.1 in icy planetesimals without.
	\end{enumerate}	
	\end{itemize}	
	\end{enumerate}

The snow line computed in our study is the distance from the central star beyond which gaseous water molecules condense as ice during the cooling of the stellar nebula (e.g. Cassen 1994; Lecar et al. 2006; Bell 2010). 
The process of condensation of gas of water molecules induces a dichotomy in the disc with the creation of a boundary beyond which abundance of solids (ices + rocks) available for planetary accretion and growth increases significantly (e.g. Stevenson \& Lunine 1988; Encrenaz 2008).
Models of protoplanetary discs by Sasselov \& Lecar (2000) around T Tauri stars result in snowlines as close as 1 AU to the central stars, depending on the stellar luminosity and the rate of accretional heating within the disc (Alibert et al. 2010). Our results show that the position of the water ice line varies from about 1 AU to 5 AU as a function of the initial surface density $\Sigma_0$ at 5.2 AU. 
When the gaseous water molecules condense, the solid surface density $\Sigma$ of the disc is assumed to increase by a factor of ~3-4 (Lecar et al. 2006; Encrenaz 2008) in theoretical studies. This assumption helps gas giant planets to form rapidly before the quick dissipation of the gas in the disc on the timescale of a few million years due to accretion onto the central protostar and evaporation from the protostar's irradiation. Our results show that the ice/rock mass ratio is of the order of magnitude 1$\pm$0.5 or lower, following the assumption on the formation of refractory organics in the disc, leading to increase only the solid surface density $\Sigma$ by approximately 1.5-2.5 beyond the ice line. As discussed in this paper, the ice/rock mass ratio is at best similar to the one discovered for the 1P/Halley (McDonnell et al. 1987) and given by Lodders (2003) for solar system and solar photospheric compositions. This increase in the amount of material free to form icy bodies and gas giant planets beyond the ice line is lower by a factor of 1.5 - 3 compared to the increase indicated by papers of planet formation (Hayashi 1981; Stevenson \& Lunine 1988; Encrenaz 2008). As the solid surface density $\Sigma$ of the disc increases only at maximum by a factor of 2 beyond the snow line, the timescale needed to quickly form giant planets should increase.

The good agreement of the ice/rock mass ratio ($\approx$1$\pm$0.5) in icy planetesimals without refractory organics with the ice-to-dust mass ratio indicated for comets (mass ratio ice/rock $\approx$ 1) and values prescribed by the Greenberg (1982) interstellar dust model, suggests that planetesimals and comets contain small fractions of refractory organics formed before the condensation of gas species in the cooling disc. In this case, planetesimals and planets formed from these small bodies (see paper 3) present high C:O depletion (molar ratio of about 0.2$\pm$0.1) relative to the initial ISM value (in our study, the solar value is 0.5) with deviations higher than 40\% (see paper 3).

The physico-chemical evolution of volatile species and grains in our model is known to be incomplete: our model does not include chemical reactions in the gas phase of the disc, gas-grain interactions and physical changes that could occur during the post-cooling phase are not taken into account. Moreover, the radial and vertical mixing of species and grains are ignored. Note also that we study only eight volatile molecules (the major components in comets) and that their abundances derive from average values observed in the ISM, except for CO for which rich and poor models were tested. Extreme lower or higher abundance of one or more species could change the chemical composition given in this study. All these assumptions may affect the chemical composition of planetesimals in the disc.

Moreover, our results suggest that all the ISM material entering the disc in the area of 30 AU around the star sublimates. However, previous works suggest that H$_2$O ice could never evaporate during the collapse of the molecular cloud beyond 12 - 30 AU (Kouchi et al. 1994; Mousis et al. 2000; Visser et al. 2011). Visser et al. (2011) show that about 60\% of the mass of H$_2$O in the disc (assuming that the abundance of H$_2$O is homogeneous in the disc) never evaporates. This means that the amount of H$_2$O remains constant without evaporation from the surface of the disc for a large part (60\%) of this molecule. In this case, the ice/rock mass ratio and chemical composition of planetesimals could change slightly compared to our results with an increase in the ice/rock ratio up to $\approx$2 in the outer part of the disc. Note also that the amorphous structure\footnote{The amorphous water ice structure can incorporate volatile molecules from the surrounding gas phase during its formation in the ISM or in the solar nebula.} could have been preserved in some part of the disc where the temperature has never exceded 110-120~K (Kouchi et al. 1994; Chick \& Cassen 1997). Many studies (Bar-Nun et al. 1987, 1988, 2007; Espinasse et al. 1991; Blake et al. 1991; Kouchi et al. 1994; Huebner et al. 2006, 2008; Notesco et al. 2003; Marboeuf et al. 2012) also suggest that comets in the solar system could be mainly composed of this structure of water ice. In this case, the volatile species initially encapsulated by the amorphous structure (Bar-Nun et al. 1987, 1988, 2007; Blake et al. 1991; Notesco et al. 2003; Yokochi et al. 2012) could slighly change the chemical composition of planetesimals computed in the outer part of the disc in this study.  

In order to better constrain the composition of planetesimals, future works should take into account the possibility of having amorphous water ice in the disc with encapsulated volatile species, and should extend the model to non-solar composition, and include the radial drift of planetesimals. 

\begin{acknowledgements}
This work has been supported by the Swiss National Science Foundation, the Center for Space and Habitibility of the University of Bern and the European Research Council under grant 239605.
\end{acknowledgements}


\begin{thebibliography}{}

\bibitem[Aikawa et al.(1999)]{1999ApJ...519..705A} Aikawa, Y., Umebayashi, T., Nakano, T., \& Miyama, S.~M.\ 1999, \apj, 519, 705 

\bibitem[Alibert et al.(2013)]{2013A&A...558A.109A} Alibert, Y., Carron, F., Fortier, A., et al.\ 2013, \aap, 558, A109 


\bibitem[Alibert et al.(2010)]{2010AsBio..10...19A} Alibert, Y., Broeg, C., Benz, W., et al.\ 2010, Astrobiology, 10, 19 

\bibitem[Alibert et al.(2005)]{2005A&A...434..343A} Alibert, Y., Mordasini, C., Benz, W., \& Winisdoerffer, C.\ 2005, \aap, 434, 343 

\bibitem[Anderson (2003)]{} Anderson, G.K.\ 2003, J. Chem. Thermodynamics, 35, 1171

\bibitem[Anderson (2007)]{} Anderson, B.J.\ 2007, Fluid Phase Equilibria, 254, 144

\bibitem[Aresu et al.(2012)]{2012A&A...547A..69A} Aresu, G., Meijerink, R., Kamp, I., et al.\ 2012, \aap, 547, A69

\bibitem[Bar-Nun et al.(2007)]{2007Icar..190..655B} Bar-Nun, A., Notesco, G., \& Owen, T.\ 2007, \icarus, 190, 655

\bibitem[Bar-Nun et al.(1988)]{1988PhRvB..38.7749B} Bar-Nun, A., Kleinfeld, I., \& Kochavi, E.\ 1988, \prb, 38, 7749 

\bibitem[Bar-Nun et al.(1987)]{1987PhRvB..35.2427B} Bar-Nun, A., Dror, J., Kochavi, E., \& Laufer, D.\ 1987, \prb, 35, 2427

\bibitem[Bell(2010)]{2010HiA....15...29B} Bell, J.~F.\ 2010, Highlights of Astronomy, 15, 29 

\bibitem[Bennett et al.(2009)]{2009ApJS..182....1B} Bennett, C.~J., Jamieson, C.~S., \& Kaiser, R.~I.\ 2009, \apjs, 182, 1

\bibitem[Bergin et al.(2007)]{2007prpl.conf..751B} Bergin, E.~A., Aikawa, Y., Blake, G.~A., \& van Dishoeck, E.~F.\ 2007, Protostars and Planets V, 751

\bibitem[Bergin et al.(1995)]{1995ApJ...441..222B} Bergin, E.~A., Langer, W.~D., \& Goldsmith, P.~F.\ 1995, \apj, 441, 222

\bibitem[Bertie \& Refaat Shehata(1984)]{1984JChPh..81...27B} Bertie, J.~E., \& Refaat Shehata, M.\ 1984, \jcp, 81, 27 

\bibitem[Bisschop et al.(2006)]{2006A&A...449.1297B} Bisschop, S.~E., Fraser, H.~J., {\"O}berg, K.~I., van Dishoeck, E.~F., \& Schlemmer, S.\ 2006, \aap, 449, 1297

\bibitem[Blake et al.(1991)]{1991Sci...254..548B} Blake, D., Allamandola, L., Sandford, S., Hudgins, D., \& Freund, F.\ 1991, Science, 254, 548 

\bibitem[Bockel{\'e}e-Morvan et al.(2004)]{2004come.book..391B} Bockel{\'e}e-Morvan, D., Crovisier, J., Mumma, M.~J., \& Weaver, H.~A.\ 2004, Comets II, 391

\bibitem[Boogert \& Ehrenfreund(2004)]{2004ASPC..309..547B} Boogert, A.~C.~A., \& Ehrenfreund, P.\ 2004, Astrophysics of Dust, 309, 547 

\bibitem[Boogert et al.(2008)]{2008ApJ...678..985B} Boogert, A.~C.~A., Pontoppidan, K.~M., Knez, C., et al.\ 2008, \apj, 678, 985 
 
\bibitem[Boogert et al.(2011)]{2011ApJ...729...92B} Boogert, A.~C.~A., Huard, T.~L., Cook, A.~M., et al.\ 2011, \apj, 729, 92 
 
\bibitem[Boonman et al.(2003)]{2003A&A...399.1063B} Boonman, A.~M.~S., van Dishoeck, E.~F., Lahuis, F., \& Doty, S.~D.\ 2003, \aap, 399, 1063  
 
\bibitem[Cassen(1994)]{1994Icar..112..405C} Cassen, P.\ 1994, \icarus, 112, 405 

\bibitem[Charnley \& Rodgers(2009)]{2009aogs...15..211C} Charnley, S.~B., \& Rodgers, S.~D.\ 2009, Advances in Geosciences, Volume 15: Planetary Science (PS), 15, 211  
 
\bibitem[Chazallon \& Kuhs 2002]{} Chazallon, B., \& Kuhs, W.F. \ 2002, J. Chem. Phys., 117, 308
 
\bibitem[Chick \& Cassen(1997)]{1997ApJ...477..398C} Chick, K.~M., \& Cassen, P.\ 1997, \apj, 477, 398 
 
\bibitem[Choukroun et al.(2006)]{2006LPI....37.1640C} Choukroun, M., Tobie, G., \& Grasset, O.\ 2006, 37th Annual Lunar and Planetary Science Conference, 37, 1640 
 
\bibitem[Choukroun et al.(2010)]{2010Icar..205..581C} Choukroun, M., Grasset, O., Tobie, G., \& Sotin, C.\ 2010, \icarus, 205, 581 
 
\bibitem[Ciesla \& Charnley(2006)]{2006mess.book..209C} Ciesla, F.~J., \& Charnley, S.~B.\ 2006, Meteorites and the Early Solar System II, 209
 
\bibitem[Cronin et al.(1993)]{1993GeCoA..57.4745C} Cronin, J.~R., Pizzarello, S., Epstein, S., \& Krishnamurthy, R.~V.\ 1993, \gca, 57, 4745  

\bibitem[Crovisier(2006)]{2006IAUS..229..133C} Crovisier, J.\ 2006, Asteroids, Comets, Meteors, 229, 133 

\bibitem[D'Alessio et al.(2001)]{2001ApJ...553..321D} D'Alessio, P., Calvet, N., \& Hartmann, L.\ 2001, \apj, 553, 321

\bibitem[Dartois et al.(1998)]{1998A&A...331..651D} Dartois, E., D'Hendecourt, L., Boulanger, F., et al.\ 1998, \aap, 331, 651 

\bibitem[Dartois et al.(1999)]{1999A&A...351.1066D} Dartois, E., Demyk, K., d'Hendecourt, L., \& Ehrenfreund, P.\ 1999, \aap, 351, 1066 

\bibitem[Dartois(2005)]{2005SSRv..119..293D} Dartois, E.\ 2005, \ssr, 119, 293 

\bibitem[Dartois(2009)]{2009ASPC..414..411D} Dartois, E.\ 2009, Cosmic Dust - Near and Far, 414, 411

\bibitem[Dartois(2011)]{2011Icar..212..950D} Dartois, E.\ 2011, \icarus, 212, 950 

\bibitem[Dauphas(2003)]{2003Icar..165..326D} Dauphas, N.\ 2003, \icarus, 165, 326 

\bibitem[Davidson et al.(1987)]{1987Natur.328..418D} Davidson, D.~W., Desando, M.~A., Gough, S.~R., Handa, Y.~P., Ratcliffe, C.~I., Ripmeester, 
J.~A., \& Tse, J.~S.\ 1987, \nat, 328, 418 


\bibitem[Doty et al.(2002)]{2002A&A...389..446D} Doty, S.~D., van Dishoeck, E.~F., van der Tak, F.~F.~S., \& Boonman, A.~M.~S.\ 2002, \aap, 389, 446 

\bibitem[Dullemond \& Dominik(2004)]{2004A&A...417..159D} Dullemond, C.~P., \& Dominik, C.\ 2004, \aap, 417, 159

\bibitem[Encrenaz(2008)]{2008ARA&A..46...57E} Encrenaz, T.\ 2008, \araa, 46, 57 

\bibitem[Espinasse et al.(1991)]{1991Icar...92..350E} Espinasse, S., Klinger, J., Ritz, C., \& Schmitt, B.\ 1991, \icarus, 92, 350 

\bibitem[Falenty et al.(2011)]{} Falenty, A., Genov, G., Hansen, T.C., Kuhs, W.H., \& Salamatin, A.N.\ 2011, J. Phys. Chem. C, 115, 4022

\bibitem[Fortes \& Choukroun(2010)]{2010SSRv..153..185F} Fortes, A.~D., \& Choukroun, M.\ 2010, \ssr, 153, 185

\bibitem[Fouchet et al.(2012)]{2012A&A...540A.107F} Fouchet, L., Alibert, Y., Mordasini, C., \& Benz, W.\ 2012, \aap, 540, A107

\bibitem[Fray \& Schmitt(2009)]{2009P&SS...57.2053F} Fray, N., \& Schmitt, B.\ 2009, \planss, 57, 2053 

\bibitem[Fray et al.(2010)]{} Fray, N., Marboeuf, U., Brissaud, O., Schmitt, B.\ 2010, J. Chem. Eng. Data, 55, 5101

\bibitem[Gabitto \& Tsouris(2010)]{} Gabitto, J.F., \& Tsouris, C. \ 2010, J. of Thermodynamics, Article ID 271291

\bibitem[Gautier et al.(2001)]{2001ApJ...550L.227G} Gautier, D., Hersant, F., Mousis, O., \& Lunine, J.~I.\ 2001a, \apjl, 550, L227 

\bibitem[Gautier et al.(2001)]{2001ApJ...559L.183G} Gautier, D., Hersant, F., Mousis, O., \& Lunine, J.~I.\ 2001b, \apjl, 559, L183 

\bibitem[Genov \& Kuhs(2003)]{2003mars.conf.3098G} Genov, G., \& Kuhs, W.~F.\ 2003, Sixth International Conference on Mars, 3098 

\bibitem[Gibb et al.(2000)]{2000ApJ...536..347G} Gibb, E.~L., Whittet, D.~C.~B., Schutte, W.~A., et al.\ 2000, \apj, 536, 347 

\bibitem[Gibb et al.(2004)]{2004ApJS..151...35G} Gibb, E.~L., Whittet, D.~C.~B., Boogert, A.~C.~A., \& Tielens, A.~G.~G.~M.\ 2004, \apjs, 151, 35 

\bibitem[Greenberg et al.(1995)]{1995ApJ...455L.177G} Greenberg, J.~M., Li, A., Mendoza-Gomez, C.~X., et al.\ 1995, \apjl, 455, L177

\bibitem[Greenberg \& D'Hendecourt(1985)]{1985iss..work..185G} Greenberg, J.~M., \& D'Hendecourt, L.~B.\ 1985, NATO ASIC Proc.~156: Ices in the Solar System, 185 

\bibitem[Greenberg(1982)]{1982come.coll..131G} Greenberg, J.~M.\ 1982, IAU Colloq.~61: Comet Discoveries, Statistics, and Observational Selection, 131

\bibitem[Grillmair et al.(2008)]{2008Natur.456..767G} Grillmair, C.~J., Burrows, A., Charbonneau, D., et al.\ 2008, \nat, 456, 767

\bibitem[Handa(1986)]{} Handa, Y.~P.\ 1986a, J. Chem. Thermodynamics, 18, 891

\bibitem[Handa(1986)]{} Handa, Y.~P.\ 1986b, J. Chem. Thermodynamics, 18, 915

\bibitem[Hayashi(1981)]{1981PThPS..70...35H} Hayashi, C.\ 1981, Progress of Theoretical Physics Supplement, 70, 35 

\bibitem[Hersant et al.(2004)]{2004P&SS...52..623H} Hersant, F., Gautier, D., \& Lunine, J.~I.\ 2004, \planss, 52, 623 

\bibitem[Huebner(2008)]{2008SSRv..138....5H} Huebner, W.~F.\ 2008, \ssr, 138, 5 

\bibitem[Huebner et al.(2006)]{2006hgdc.conf.....H} Huebner, W.~F., Benkhoff, J., Capria, M.-T., Coradini, A., de Sanctis, C., Orosei, R., 
\& Prialnik, D.\ 2006, Published for The International Space Science Institute, Bern, Switzerland, by ESA Publications Division, Noordwijk, The Netherlands.

\bibitem[Hueso \& Guillot(2005)]{2005A&A...442..703H} Hueso, R., \& Guillot, T.\ 2005, \aap, 442, 703 

\bibitem[Ipatov \& Mather(2006)]{2006AdSpR..37..126I} Ipatov, S.~I., \& Mather, J.~C.\ 2006, Advances in Space Research, 37, 126


\bibitem[Ipatov \& Mather(2007)]{2007IAUS..236...55I} Ipatov, S.~I., \& Mather, J.~C.\ 2007, IAU Symposium, 236, 55

\bibitem[Iro et al.(2003)]{2003Icar..161..511I} Iro, N., Gautier, D., Hersant, F., Bockel{\'e}e-Morvan, D., \& Lunine, J.~I.\ 2003, \icarus, 161, 511 

\bibitem[Irvine et al.(2000)]{2000prpl.conf.1159I} Irvine, W.~M., Schloerb, F.~P., Crovisier, J., Fegley, B., Jr., \& Mumma, M.~J.\ 2000, Protostars and Planets IV, 1159

\bibitem[Kang et al.(2001)]{} Kang, S.P., Lee, H., Ryu, B.-J.\ 2001, J. Chem. Thermodynamics, 33, 513

\bibitem[Kargel(1998)]{1998ASSL..227....3K} Kargel, J.~S.\ 1998, Solar System Ices, 227, 3 

\bibitem[Kissel et al.(1986)]{1986Natur.321..280K} Kissel, J., Sagdeev, R.~Z., Bertaux, J.~L., et al.\ 1986a, \nat, 321, 280 


\bibitem[Kissel et al.(1986)]{1986Natur.321..336K} Kissel, J., Brownlee, D.~E., Buchler, K., et al.\ 1986b, \nat, 321, 336 

\bibitem[Kouchi et al.(2002)]{2002ApJ...566L.121K} Kouchi, A., Kudo, T., Nakano, H., et al.\ 2002, \apjl, 566, L121 

\bibitem[Kouchi et al.(1994)]{1994A&A...290.1009K} Kouchi, A., Yamamoto, T., Kozasa, T., Kuroda, T., \& Greenberg, J.~M.\ 1994, \aap, 290, 1009 

\bibitem[Kuhs et al.(2006)]{} Kuhs, W.F., Staykova, D.K., Salamatin, A.N.\ 2006, J. Phys. Chem. B, 110, 13283

\bibitem[Jessberger et al.(1988)]{1988Natur.332..691J} Jessberger, E.~K., Christoforidis, A., \& Kissel, J.\ 1988, \nat, 332, 691

\bibitem[Lahuis et al.(2007)]{2007ApJ...665..492L} Lahuis, F., van Dishoeck, E.~F., Blake, G.~A., et al.\ 2007, \apj, 665, 492

\bibitem[Langer et al.(2000)]{2000prpl.conf...29L} Langer, W.~D., van Dishoeck, E.~F., Bergin, E.~A., et al.\ 2000, Protostars and Planets IV, 29

\bibitem[Lecar et al.(2006)]{2006ApJ...640.1115L} Lecar, M., Podolak, M., Sasselov, D., \& Chiang, E.\ 2006, \apj, 640, 1115 

\bibitem[Lee et al.(2008)]{2008M&PS...43.1351L} Lee, J.-E., Bergin, E.~A., \& Lyons, J.~R.\ 2008, Meteoritics and Planetary Science, 43, 1351 

\bibitem[Lewis \& Prinn(1980)]{1980ApJ...238..357L} Lewis, J.~S., \& Prinn, R.~G.\ 1980, \apj, 238, 357 

\bibitem[Lewis(1972)]{1972Icar...16..241L} Lewis, J.~S.\ 1972, \icarus, 16, 241 

\bibitem[Lodders(2003)]{2003ApJ...591.1220L} Lodders, K.\ 2003, \apj, 591, 1220 

\bibitem[Lunine \& Stevenson(1987)]{1987Icar...70...61L} Lunine, J.~I., \& Stevenson, D.~J.\ 1987, \icarus, 70, 61 

\bibitem[Marboeuf et al.(2014)]{} Marboeuf, U., Thiabaud, A., Alibert A., Cabral, N., \& Benz, W.\ 2014, \aap, submitted

\bibitem[Marboeuf et al.(2012)]{2012A&A...542A..82M} Marboeuf, U., Schmitt, B., Petit, J.-M., Mousis, O., \& Fray, N.\ 2012, \aap, 542, A82

\bibitem[Marboeuf et al.(2008)]{2008ApJ...681.1624M} Marboeuf, U., Mousis, O., Ehrenreich, D., Alibert, Y., Cassan, A., Wakelam, V., \& Beaulieu, J.-P.\ 2008, \apj, 681, 1624 

\bibitem[McDonnell et al.(1987)]{1987A&A...187..719M} McDonnell, J.~A.~M., et al.\ 1987, \aap, 187, 719

\bibitem[Meijerink et al.(2012)]{2012A&A...547A..68M} Meijerink, R., Aresu, G., Kamp, I., et al.\ 2012, \aap, 547, A68

\bibitem[Miller(1961)]{1961PNAS...47.1798M} Miller, S.~L.\ 1961, Proceedings of the National Academy of Science, 47, 1798

\bibitem[Mohammadi \& Richon(2010)]{} Mohammadi, A.H., \& Richon, D.\ 2010, Ind. Eng. Chem. Res., 49, 3976 

\bibitem[Moore et al.(2007)]{2007Icar..190..260M} Moore, M.~H., Ferrante, R.~F., Hudson, R.~L., \& Stone, J.~N.\ 2007, \icarus, 190, 260

\bibitem[Mousis et al.(2010)]{2010FaDi..147..509M} Mousis, O., Lunine, J.~I., Picaud, S., \& Cordier, D.\ 2010, Faraday Discussions, 147, 509

\bibitem[Mousis \& Gautier(2004)]{2004P&SS...52..361M} Mousis, O., \& Gautier, D.\ 2004, \planss, 52, 361 

\bibitem[Mousis et al.(2002)]{2002Icar..156..162M} Mousis, O., Gautier, D., \& Bockel{\'e}e-Morvan, D.\ 2002, \icarus, 156, 162 

\bibitem[Mousis et al.(2000)]{2000Icar..148..513M} Mousis, O., Gautier, D., Bockel{\'e}e-Morvan, D., Robert, F., Dubrulle, B., \& Drouart, A.\ 2000, \icarus, 148, 513

\bibitem[Mumma \& Charnley(2011)]{2011ARA&A..49..471M} Mumma, M.~J., \& Charnley, S.~B.\ 2011, \araa, 49, 471 

\bibitem[Mumma(1997)]{1997ASPC..122..369M} Mumma, M.~J.\ 1997, From Stardust to Planetesimals, 122, 369

\bibitem[Neufeld \& Hollenbach(1994)]{1994ApJ...428..170N} Neufeld, D.~A., \& Hollenbach, D.~J.\ 1994, \apj, 428, 170

\bibitem[Nomura \& Millar(2004)]{2004A&A...414..409N} Nomura, H., \& Millar, T.~J.\ 2004, \aap, 414, 409

\bibitem[Notesco et al.(2003)]{2003Icar..162..183N} Notesco, G., Bar-Nun, A., \& Owen, T.\ 2003, \icarus, 162, 183

\bibitem[Notesco \& Bar-Nun(2000)]{2000Icar..148..456N} Notesco, G., \& Bar-Nun, A.\ 2000, \icarus, 148, 456 

\bibitem[{\"O}berg et al.(2008)]{2008ApJ...678.1032O} {\"O}berg, K.~I., Boogert, A.~C.~A., Pontoppidan, K.~M., et al.\ 2008, \apj, 678, 1032 

\bibitem[Parrish \& Prausnitz(1972)]{} Parrish, W.R., Prausnit, J.M.\ 1972, Ind. Eng. Chem. Process Des. Develop., 11, 26

\bibitem[Pollack et al.(1994)]{1994ApJ...421..615P} Pollack, J.~B., Hollenbach, D., Beckwith, S., et al.\ 1994, \apj, 421, 615

\bibitem[Pontoppidan et al.(2003)]{2003A&A...404L..17P} Pontoppidan, K.~M., Dartois, E., van Dishoeck, E.~F., Thi, W.-F., \& d'Hendecourt, L.\ 2003, \aap, 404, L17 

\bibitem[Pontoppidan et al.(2008)]{2008ApJ...678.1005P} Pontoppidan, K.~M., Boogert, A.~C.~A., Fraser, H.~J., et al.\ 2008, \apj, 678, 1005

\bibitem[Prinn \& Fegley(1981)]{1981ApJ...249..308P} Prinn, R.~G., \& Fegley, B., Jr.\ 1981, \apj, 249, 308 

\bibitem[Prinn \& Fegley(1989)]{1989oeps.book...78P} Prinn, R.~G.~P., \& Fegley, B., Jr.\ 1989, Origin and Evolution of Planetary and Satellite Atmospheres, 78 

\bibitem[Rydzy et al.(2007)]{} Rydzy, M.B., Schicks, J.M., Naumann, R., Erzinger, J.\ 2007, J. Phys. Chem. B, 111, 9539

\bibitem[Sasselov \& Lecar(2000)]{2000ApJ...528..995S} Sasselov, D.~D., \& Lecar, M.\ 2000, \apj, 528, 995 

\bibitem[Schmitt(1986)]{} Schmitt, B.\ 1986, PhD Thesis, Univ. Joseph Fourier, Grenoble, France

\bibitem[Schutte \& Khanna(2003)]{2003A&A...398.1049S} Schutte, W.~A., \& Khanna, R.~K.\ 2003, \aap, 398, 1049 

\bibitem[Sekine et al.(2005)]{2005Icar..178..154S} Sekine, Y., Sugita, S., Shido, T., et al.\ 2005, \icarus, 178, 154 

\bibitem[Shakura \& Sunyaev(1973)]{1973A&A....24..337S} Shakura, N.~I., \& Sunyaev, R.~A.\ 1973, \aap, 24, 337 

\bibitem[Shin et al.(2012)]{} Shin, K., Kumar, R., Udachin, K.A., Alavi, S., Ripmeester, J.A., \ 2012, PNAS 

\bibitem[Sloan \& Koh(2008)]{} Sloan, E. D.; Koh, C. A. \ 2008, Clathrate Hydrates of Natural Gases, third
ed.; CRC Press, Taylor \& Francis Group: Boca Raton

\bibitem[Staykova(2004)]{} Staykova, D. K.\ 2004, Ph.D. Thesis, Universitat Gottingen.

\bibitem[Staykova et al.(2003)]{} Staykova, D. K., Kuhs, W. F., Salamatin, A. N., \& Hansen, T.\ 2003, J. Phys. Chem. B, 107, 10299

\bibitem[Stevenson \& Lunine(1988)]{1988Icar...75..146S} Stevenson, D.~J., \& Lunine, J.~I.\ 1988, \icarus, 75, 146 

\bibitem[Sun \& Duan(2005)]{} Sun, R., \& Duan, Z.\ 2005, Geochimica et Cosmochimica Acta, 69, 4411

\bibitem[Sun \& Mohanty(2006)]{} Sun, X., \& Mohanty, K.K.\ 2006, Chemical Engineering Science, 61, 3476 

\bibitem[Talbi \& Herbst(2002)]{2002A&A...386.1139T} Talbi, D., \& Herbst, E.\ 2002, \aap, 386, 1139

\bibitem[Tancredi et al.(1994)]{1994A&A...286..659T} Tancredi, G., Rickman, H., \& Greenberg, J.~M.\ 1994, \aap, 286, 659

\bibitem[Thiabaud et al.(2014)]{2014A&A...562A..27T} Thiabaud, A., Marboeuf, U., Alibert, Y., et al.\ 2014, \aap, 562, A27

\bibitem[Thomas et al.(2009)]{2009P&SS...57...42T} Thomas, C., Mousis, O., Picaud, S., \& Ballenegger, V.\ 2009, \planss, 57, 42 

\bibitem[Van der Waals \& Platteeuw(1959)]{} Van der Waal, J.H., \& Platteeuw, J.C.\ 1959, Advances in Chemical Physics, interscience New$-$York

\bibitem[van Dishoeck(2004)]{2004ARA&A..42..119V} van Dishoeck, E.~F.\ 2004, \araa, 42, 119 

\bibitem[Visser et al.(2011)]{2011A&A...534A.132V} Visser, R., Doty, S.~D., \& van Dishoeck, E.~F.\ 2011, \aap, 534, A132

\bibitem[Visser \& Dullemond(2010)]{2010A&A...519A..28V} Visser, R., \& Dullemond, C.~P.\ 2010, \aap, 519, A28 

\bibitem[Visser et al.(2009)]{2009A&A...495..881V} Visser, R., van Dishoeck, E.~F., Doty, S.~D., \& Dullemond, C.~P.\ 2009, \aap, 495, 881

\bibitem[Walsh et al.(2010)]{2010ApJ...722.1607W} Walsh, C., Millar, T.~J., \& Nomura, H.\ 2010, \apj, 722, 1607

\bibitem[Wang et al.(2002)]{} Wang, X., Schultz, A. J., \& Halpern, Y.\ 2002, J. Phys. Chem. A, 106, 7304

\bibitem[White(2010)]{2010PhDT........33W} White, D.~W.\ 2010, Ph.D.~Thesis,  

\bibitem[Whittet et al.(2007)]{2007ApJ...655..332W} Whittet, D.~C.~B., Shenoy, S.~S., Bergin, E.~A., et al.\ 2007, \apj, 655, 332 

\bibitem[Willacy \& Woods(2009)]{2009ApJ...703..479W} Willacy, K., \& Woods, P.~M.\ 2009, \apj, 703, 479

\bibitem[Yokochi et al.(2012)]{2012Icar..218..760Y} Yokochi, R., Marboeuf, U., Quirico, E., \& Schmitt, B.\ 2012, \icarus, 218, 760


\end{thebibliography}
\end{document}